\documentclass [11pt]{article}
\title{New Perspectives on the Relativistically Rotating Disk\\and Non-time-orthogonal Reference Frames}
\author{Robert D. Klauber\\  \\1100 University Manor Dr, 38B\\Fairfield, Iowa 52556\\rklauber@netscape.net}
\date{October 1998}
\usepackage {graphicx}
\usepackage {longtable}

\begin{document}

\maketitle

\bigskip

\begin {abstract}


The rotating disk problem is analyzed on the premise that proper 
interpretation of experimental evidence leads to the conclusion that the 
postulates upon which relativity theory is based, particularly the 
invariance of the speed of light, are not applicable to rotating frames. 
Different postulates based on the Sagnac experiment are proposed, and from 
these postulates a new relativistic theory of rotating frames is developed 
following steps similar to those initially followed by Einstein for 
rectilinear motion. The resulting theory agrees with all experiments, 
resolves problems with the traditional approach to the rotating disk, and 
exhibits both traditionally relativistic and non-relativistic 
characteristics. Of particular note, no Lorentz contraction exists on the 
rotating disk circumference, and the disk surface, contrary to the 
assertions of Einstein and others, is found to be Riemann flat. The variable 
speed of light found in the Sagnac experiment is then shown to be 
characteristic of non-time-orthogonal reference frames, of which the 
rotating frame is one. In addition, the widely accepted postulate for the 
equivalence of inertial and non-inertial standard rods with zero relative 
velocity, used liberally in prior rotating disk analyses, is shown to be 
invalid for such frames. Further, the new theory stands alone in correctly 
predicting what was heretofore considered a "spurious" non-null effect on 
the order of 10$^{{\rm -} {\rm 1}{\rm 3}}$ found by Brillet and Hall in the 
most accurate Michelson-Morley type test to date. The presentation is simple 
and pedagogic in order to make it accessible to the non-specialist.

\bigskip


\end{abstract}

\section*{1 BACKGROUND}

\subsection*{1.1 Perspectives of Einstein and Others}

Albert Einstein never published a technical paper directly addressing the 
problem of the relativistically rotating rigid disk, although in private 
writings\cite{Stachel:1980}, in three books for the general 
audience\cite{Einstein:1920},\cite{Einstein:1921},\cite{Einstein:1938}, 
and as support for the use of generalized coordinates in his landmark 1916 
paper\cite{Einstein:1916}, he purported that the space of such a rotating 
disk is curved, not flat. He further attributed his early insights into 
general relativity theory to be a direct result of contemplating the 
curvature of such a rotating system. His perspective on the problem is 
revealed in a private 1919 letter found by Stachel\cite{Ref:1} in the 
Einstein Archives at the Institute for Advanced Study in Princeton, New 
Jersey. 

In that letter Einstein considers measuring rods laid out along the disk's 
radii and circumference and assumes Lorentz contraction exists along the 
circumference due to the tangential velocity $v $= $\omega r$ of the disk at a 
given radius $r$. In Einstein's words:

\bigskip

\textit{....imagine a "snapshot" taken from [the non-rotating frame] ... On this snapshot the radial measuring rods have the length l, the tangential ones, however, the length l}$(1 - $v$^{2}/c^{2})^{1 / 2}$\textit{. The "circumference" [therefore is] U} = [\textit{2$\pi $r}]/($1$ - v$^{2}$/c$^{2}$)$^{{\rm 1}{\rm /} {\rm 
2}} . $

\bigskip

He repeated this type of reasoning elsewhere [2, 3, 4], and used it to 
conclude that the rotating disk surface is not Euclidean since $U > 2\pi r$.

But not everyone agrees. In a 1951 letter\cite{Ref:2} Einstein noted that 
Eddington and Lorentz considered the geometry on the disk to be flat, and he 
stated that he did not know what they meant. No references recounting 
Eddington's and Lorentz's thinking on the subject seem to be available, but 
others, such as Levy\cite{Levy:1939}, appear to agree with them. 
Strauss\cite{Strauss:1974} concludes that the space is curved, but argues 
that Einstein's logic was flawed, noting that

\bigskip

\textit{If the measuring rods laid along the circumference of the rotating disk are Lorentz contracted with respect to the inertial frame, so are the distances on the circumference they are supposed to measure; hence the two effects would cancel each other, and the ratio} U/D\textit{ would turn out to equal $\pi $ as in the Euclidean plane.}

\bigskip

Gr{\o}n \cite{Rotating:1977}, however, citing M{\o}ller 
\cite{ller:1969} and Landau and Lifshitz \cite{Landau:1962}, contends 
that this argument is wrong, Weber \cite{Weber:1997} supports Gr{\o}n's 
view, and Stachel [1] effectively concedes the point to Einstein.

\subsection*{1.2 Relevant Relativity Principles}

Special relativity is restricted to inertial systems and is derived from two 
symmetry postulates:

\bigskip

1. The speed of light is the same for all inertial observers (it is 
invariant) and equals $c.$

2. There is no preferred inertial reference frame. (Velocity is relative, 
and the laws of nature are covariant, i.e., the same for all inertial 
observers.)

\bigskip

General relativity is applicable to non-inertial systems and is based on 
generalizations of the above two postulates embellished with other certain 
principles/assumptions, including:

\bigskip

1. The speed of light is invariant and equals $c$ for non-inertial observers 
provided that it is measured locally by local standard clocks and measuring 
rods \cite{This:1}.

2. There is no preferred non-inertial frame. (The laws of nature are also 
covariant for non-inertial observers, although coordinate metrics different 
from those of special relativity are needed to represent those laws.)

3. Gravity and acceleration are \textit{locally} indistinguishable, i.e., the equivalence 
principle. (Over finite distances, gravity can, however, be distinguished 
from acceleration due to the presence of gravitational tidal forces or 
geodesic deviation.)

4. Neither gravity nor acceleration changes the length of a standard rod or 
the rate of a standard clock relative to a nearby freely falling (inertial) 
standard rod or standard clock having the same velocity. (This is an 
assumption rarely emphasized in most texts, though M{\o}ller$^{{\rm} }$[11] 
makes the point clearly, and Einstein \cite{Ref:3} emphasized it a number 
of times.) We will call this the "surrogate frames postulate" or when used 
with reference to standard rods, the "surrogate rods postulate."

\bigskip

The first general relativity point above is often a source of confusion, as 
it is sometimes said that the speed of light in general relativity can be 
different than $c$. This is true if, for example, one measures the speed of 
light near a massive star using a clock based on earth. (Time on such a 
clock is effectively the coordinate time in a Schwarzchild coordinate 
system.) As is well known, due to the intense gravitation field, the passage 
of time close to the star is dilated relative to earth time, and one would 
indeed calculate a light speed other than $c$. However, use of standard rods 
and clocks adjacent the light ray itself would result in a speed of 
precisely $c$.

Other confusion exists for scenarios where spacetime itself expands or 
contracts. For example, just after the big bang, space itself was expanding 
much like the surface of a balloon being blown up. A photon in space 
(analogous to an ant on the surface of the balloon) at a different location 
than an observer could then move away from the observer faster than $c$ 
(analogous to faster than the ant can crawl on the surface) because the 
space (balloon surface) between the photon and the observer is itself 
expanding. Yet a photon spatially coincident with an observer could never be 
seen by that observer to have speed greater than $c$, and local standard rods 
and clocks adjacent any photon would find its speed equaling $c$ regardless of 
the dynamical state of spacetime itself.

\subsection*{1.3 Nomenclature and Definitions}

Upper-case letters herein shall refer to inertial systems; lower case to 
non-inertial systems. For example K shall designate the non-rotating (lab) 
frame; k, the rotating frame.

Flat spacetime will be referred to as "Minkowski space"; whereas the term 
"Minkowski metric" will be limited to refer only to a Minkowskian set of 
coordinates (Cartesian plus time) used within that flat space. Hence a 
Minkowski space need not be represented solely by a Minkowski metric, and we 
will in fact use cylindrical coordinates, i.e., (\textit{cT,R,}$\Phi $,$Z$), for the flat 
non-rotating inertial frame.

Though much of the paper may be understood with no working knowledge of 
differential geometry (the mathematics of general relativity), in Sec. 4 it 
is needed to derive certain results. These derivations are based on a 
Minkowski metric defined as

\begin{equation}
\label{eq1}
\eta _{\alpha ,\beta}  = {\left[ {{\begin{array}{*{20}c}
 { - 1} \hfill & {0} \hfill & {0} \hfill & {0} \hfill \\
 {0} \hfill & {1} \hfill & {0} \hfill & {0} \hfill \\
 {0} \hfill & {0} \hfill & {1} \hfill & {0} \hfill \\
 {0} \hfill & {0} \hfill & {0} \hfill & {1} \hfill \\
\end{array}} } \right]}\;,\quad \quad \alpha ,\beta = 0,1,2,3
\end{equation}

\bigskip

To eliminate confusion for the non-specialist, and to more readily compare 
results with those of Einstein and others, we employ cgs, not geometrized, 
units where $c = $2.998 X 10$^{{\rm 1}{\rm 0}}$ cm/sec.

\subsection*{1.4 Rotating Disk Experimental Evidence}
\label{subsec:mylabel1}

\subsubsection*{1.4.1 The speed of light.}

In 1913 Sagnac \cite{Post:1967} first demonstrated experimentally that 
rotating disks exhibit a remarkable property, the significance of which the 
present author believes has been completely overlooked ever since. That is, 
the local speed of a beam of light tangent to the disk circumference is not 
invariant, and not isotropic (as seen from the disk).

\begin{figure}
\centering
\includegraphics*[bbllx=0.26in,bblly=0.13in,bburx=3.01in,bbury=3.02in,scale=1.00]{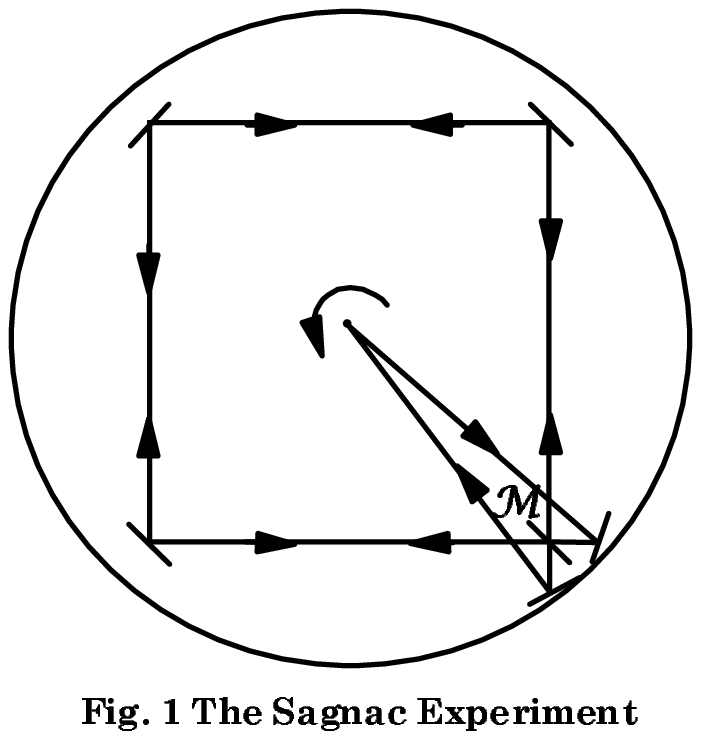}
\end{figure}

\bigskip

Fig. 1 depicts the Sagnac experiment schematically. A light beam is emitted 
radially from the center of a rotating disk and is split by a half silvered 
mirror M at radius $r$. From there one part of the beam is reflected by mirrors 
appropriately placed on the disk such that it travels in one direction 
around the circumference. The other half of the beam travels the same route 
over precisely the same distance, but in the opposite direction. The beams 
then meet up again and are reflected back to the center where interference 
of the two beams results in a fringing, i.e., a displacement of one light 
wave with respect to the other.

\bigskip

This is exactly the effect Michelson and Morley were first looking for, but 
could, due to now well known relativistic effects, never detect. Fringing 
results in either experiment would have implied different velocities of 
light in different directions. Hence, while the Michelson-Morley result 
indicated that for translational motion the speed of light is invariant and 
isotropic, the Sagnac experiment indicates that for rotational motion, no 
such conclusion may be drawn.

The results of Sagnac and others who have repeated his experiment have 
experimental accuracy only to first order in $v/c$ = $\omega r/c$, and indicate that 
the speed of a light ray tangent to the circumference measured locally on 
the disk is equal to \cite{Post:1}

\begin{equation}
\label{eq2}
\vert v_{{\rm l}{\rm i}{\rm g}{\rm h}{\rm t}} {\rm \vert} {\rm} {\rm}  \cong 
{\rm} {\rm} c{\rm} {\rm} \pm {\rm} {\rm} \omega r
\end{equation}

\noindent
where the approximately equal sign implies accuracy to first order, and the 
sign in front of the last term depends on the relative direction of the rim 
tangent and light ray velocities.

These results should, in fact, be expected. An inertial (non-rotating) 
observer of the two light rays would see each of them having the speed $c$, and 
during the time they are traveling around the circumference the disk would 
rotate some amount. Hence one ray would meet back up with the half-silvered 
mirror M before the other, and an observer fixed on the disk at M would 
conclude that the speeds in each direction were different. Further, the 
difference can be readily calculated, to first order, to be that shown in 
Eq. (\ref{eq2}). (Selleri \cite{Selleri:1997} makes the calculation rigorously to 
all orders.) Still further, the effect is \textit{local} since angular velocity is 
constant and due to symmetry, any segment of a constant radius path (in the 
ideal experimental design) is equivalent to any other segment. Hence any 
global (average) speed effect measured by Sagnac over finite times and 
distances is equal to the local (infinitesimal) speed at any point on the 
circumference.

\subsubsection*{1.4.2 Absolute nature of rotational velocity.} 
\label{subsubsec:mylabel1}

The Michelson-Morley experiment also implied that translational velocities 
are relative, and that there is no preferred system of reference (no 
"ether").

Rotational velocities, on the contrary, are absolute. (The term "absolute" 
herein implies accordance with Mach's principle, i.e., absoluteness with 
respect to the distant galaxies, within Einstein's relativistic theory of 
flat spacetime). This is due, at least in part, to the absolute nature of 
the radially directed accelerations experienced by any rotating object. 
Hence, for rotational velocities there is a preferred frame, and it is the 
one in which no radial accelerations are experienced. Any observer, in any 
frame, can tell which system is the non-rotating one, i.e., the "preferred 
frame", and how much each of the other frames is rotating relative to it. 
This can be done by watching the motion of a Foucault pendulum, by noticing 
whether or not there is a Coriolis "force", or by a number of other means.

\section*{2 DIFFICULTIES WITH THE TRADITIONAL VIEW}

\subsection*{2.1 The Postulates}

The reader has no doubt noticed that the experimental results of Sec. 1.4 
above appear to clash with the very postulates of Sec. 1.2 upon which the 
theory of relativity was founded. That is, all of the relativistic behavior 
with which we have become so familiar in the twentieth century, such as the 
Lorentz contraction and the lack of agreement on simultaneity, are a direct 
result of i) invariance of the speed of light, and ii) "reference frame 
democracy" (all frames are equal). Yet, the author contends, these 
postulates simply do not hold for rotating frames.

This point seems to have been overlooked for two reasons. Firstly, much of 
the literature [16] covering the Sagnac effect focuses on the fringe effect 
\textit{per se}, and its concomitant mathematical description, without noting the 
significant implications such fringing has for the speed of light. Secondly, 
of those who were aware that this implied a variable speed of light, most 
probably glossed over the fact by assuming that in some manner it was merely 
another general relativistic (non-inertial systems) manifestation of the 
"light speed unequal to $c$ effect". Yet, unlike the examples provided in Sec. 
1.2, there is no expansion or contraction of spacetime associated with the 
rotating disk, and the fringing implies a true difference in the \textit{local} measurement 
of light speed.

Therefore one should, indeed \textit{must}, expect rotating frame behavior to differ from 
that of translational motion. Relativistic effects such as the Lorentz 
contraction are not given \textit{a priori}; they are derived. And they are derived from 
different empirically based principles than those governing rotational 
motion. Hence, we should be wary of conclusions drawn by simply applying 
derived tenets of relativity theory to rotating disks, as Einstein and 
others have done.

\subsection*{2.2 Geodesic Deviation}

The Riemann curvature tensor (or simply "Riemann") $R$ is a measure of the 
curvature of a given space. It is defined by virtue of the geodesic 
deviation\textit{} equation of differential geometry \cite{Misner:1973},

\begin{equation}
\label{eq3}
\nabla _{u} \nabla _{u} n\,\;\, + \;\;R(...,u,n,u)\; = \;{\rm {\bf 0}}\,.
\end{equation}

The first term above represents the \textit{deviation} between \textit{two} geodesics, i.e., the rate of 
change of the rate of change in proper distance of the perpendicular from 
the first geodesic to the second as one travels along the first. If $R$ is 
zero, then Eq. (\ref{eq3}) dictates that every pair of geodesics which are initially 
parallel will stay parallel along their entire length. The proper distance 
between them will never change, and they will never intersect. Since this is 
true only of flat spaces, a zero value for $R$ means the space is flat. If$R 
\ne 0$, such as on the surface of a globe, two geodesics (e.g., lines of 
longitude) which start out parallel (at the equator), don't stay parallel 
(and cross at the poles). $R$ is characteristic of the space itself, not the 
coordinate system used within that space. If, for example, $R$ = 0 for a 2D 
flat space with Cartesian coordinate system, when we transform to a polar 
coordinate system, we still have $R$ = 0.

It is commonly known \cite{Ref:4,Ref:5} that the four-dimensional 
(4D) spacetime of the rotating system (denoted k) is Riemann flat since $R$ = 0 
in the non-rotating frame (denoted K), and the rotating system coordinates 
are obtained by simply transforming the 4D coordinates of K into k. If 
Riemann is zero in K, it must also be zero in k. However, it does not 
necessarily follow that the subspace of the disk surface, embedded in the 4D 
space, is flat. By analogy, the 2D subspace surface of a sphere embedded in 
a flat 3D space is not itself flat.

Particles attached to the rotating disk undergo acceleration and hence do 
not follow geodesic paths. The path of a free particle or light ray, 
however, is a geodesic, and though it is straight as seen from K, it looks 
curved, even "corkscrew-like," as seen from k.

The question of flatness for the subspace of the disk can be addressed by 
considering two free particles traveling at the same velocity in K in the 
plane of the disk surface (i.e., the axial coordinate $Z$ = constant), and 
tracing out parallel lines in K. The observer in K sees them as straight and 
never intersecting. The rotating observer sees them as corkscrew-like and 
never intersecting. The point is that the geodesic equation, from which the 
Riemann tensor is defined, relates to the "never intersecting" part, not the 
"non-straight" part. In a curved space there is geodesic \textit{deviation}. It says two 
geodesics \textit{deviate} in their behavior. The two geodesics in question do not. Further, 
the two geodesics travel \textit{in the plane of the disk} \textit{surface.} Regardless of how one wishes to define the disk 
surface and all of the issues of simultaneity involved (see later sections), 
the basic fact remains that for $z$ = constant, the two geodesics do not 
deviate (i.e., they never cross). For a Riemann curved surface they must 
deviate. Therefore the disk surface is Riemann flat.

By analogy, two parallel geodesics which appear straight in an inertial 
system appear curved to an observer in a rectilinearly accelerating system. 
But they never appear to cross to the accelerating observer, and the proper 
distance between them never changes. As is well known, the space of a 
rectilinearly accelerating system is flat \cite{Ref:6}. This is in full 
accord with Eq. (\ref{eq3}) since geodesic deviation for such a system is zero, and 
so is Riemann, even though geodesics themselves do not look straight.

Geodesic deviation causes tidal forces, the stretching and compressing of a 
finite sized object in free fall (i.e., traveling a geodesic). Gravity tries 
to make one side of the object accelerate in a different direction, or at a 
different rate, than the other side. But a finite sized object traveling 
along a geodesic in the plane of the rotating disk would not$^{{\rm 
}}$experience any tidal forces, and \textit{all} observers, whether on the disk, the 
lab, or anywhere else, would agree there is no stress or strain within the 
object. Hence Riemann is zero along the path of the object, and the surface 
of the disk can not be curved.

\subsection*{2.3 Tangent Frames and the Discontinuity in Time}
\label{subsec:mylabel2}

Applying traditional relativistic concepts directly to the rotating disk 
leads to another striking difficulty. It predicts a discontinuity in time on 
the surface of the disk, and in addition, the location of that discontinuity 
is arbitrary, being merely a function of the particular predilections of the 
observer. In other words, a continuous standard tape measure extending one 
circumference around the rim would not meet back up with itself at the same 
point in time. The logic leading to this conclusion follows.

In order to evaluate disk curvature, prior researchers have invoked the 
"surrogate frames postulate" (see general relativity principle 4 of Sec. 
1.2) and used a series of inertial reference frames tangent to the disk rim 
with velocities equal to that of the rim edge (i.e., with $v $= $\omega r$). It 
is argued that since acceleration body forces do not affect standard rod 
length, rods in these inertial frames should be affected in precisely the 
same manner as rods aligned with, and attached to, the edge of the disk rim 
in k.

The problems with this approach can be illustrated with the aid of Fig. 2. 
Inertial measuring rods in inertial frames K$_{{\rm 1}}$ to K$_{{\rm 8}}$ 
with speeds $\omega r$ can be imagined as shown. For practical reasons we 
only show eight finite length rods, and we consider them as a symbolic 
representation of an infinite number of rods of infinitesimal length. A and 
B are \textit{events} located in space at the endpoints of the K$_{{\rm 1}}$ rod which are 
simultaneous as seen from K$_{{\rm 1}}$; B and C are events located in space 
at the endpoints of the K$_{{\rm 2}}$ rod which are simultaneous in K$_{{\rm 
2}}$; and so on for the other events C to J. A,B, ...J can be envisioned as 
flashes of light emitted by bulbs situated equidistantly around the disk 
rim.

\begin{figure}
\centering
\includegraphics*[bbllx=0.26in,bblly=0.13in,bburx=5.25in,bbury=2.47in,scale=1.00]{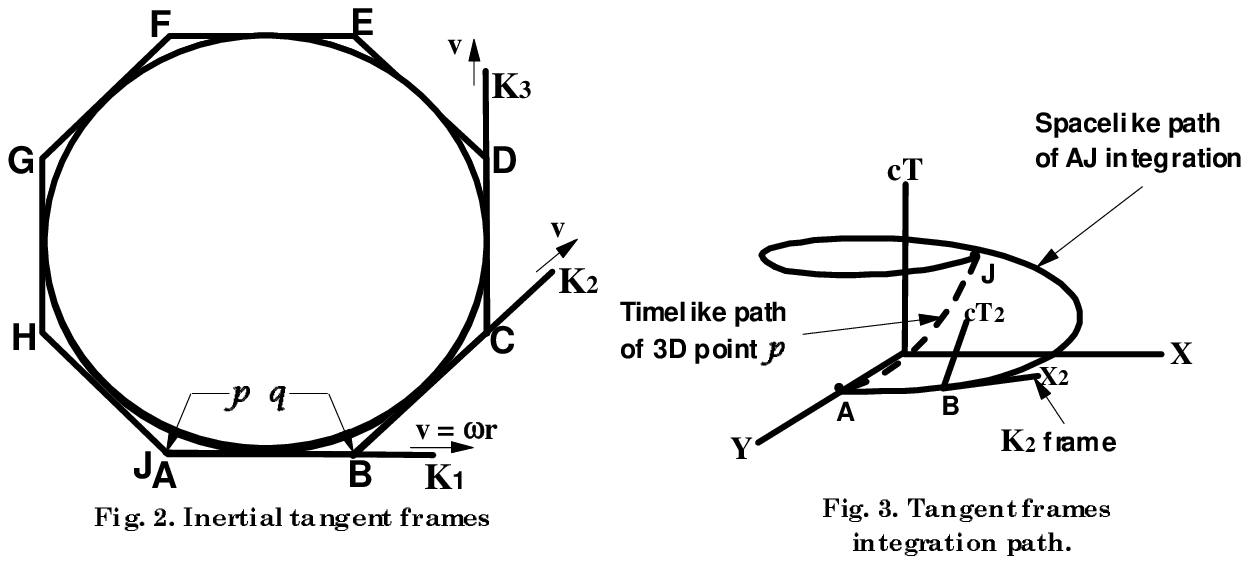}
\end{figure}

\bigskip

\noindent
p is a spatial (three dimensional) point fixed to the disk at which both A 
and J occur. q is the spatial point on the disk at which B occurs. In 
principle, A, B, ... J, as well as p and q are located on the disk rim 
though they may not look so in Fig. 2 since the tangent rods shown are not 
infinitesimal in length.

Note that although events A and B are simultaneous as seen from K$_{{\rm 
1}}$, they are not simultaneous as seen in K (via standard relativity theory 
for two inertial frames in relative motion). As seen from K, A occurs before 
B. Similarly, B occurs before C, and so on around the rim. If the events are 
light flashes, a ground based observer looking down on the disk would see 
the A flash, then B, then C, etc. Hence we conclude that as seen from K, A 
occurs before J even though A and J are both located at the same 3D point p 
fixed to the rim. As seen from K, during the time interval between events A 
and J the disk rotates, and hence the point p moves. (As an aside, Fig. 2 
can now be seen to be merely symbolic since events A to J would not in 
actuality be seen from K to occur at the locations shown in Fig. 2. That is, 
by the time the K observer sees the B flash, the disk has rotated a little. 
It rotates a little more before he sees the C flash, etc.) 

According to the traditional treatment of the rotating disk, one then uses 
the K$_{{\rm i}}$ rods and integrates (adds the rod lengths) along the path 
AB ...J, moving sequentially from tangent inertial frame to tangent inertial 
frame. This path is represented by the solid line in Fig. 3, and one can 
visualize small Minkowski coordinate frames situated at every point along 
the curve AJ (see K$_{{\rm 2}}$ in Fig. 3) with integration taking place 
along a series of spatial axes (such as X$_{{\rm 2}}$ in Fig. 3). By doing 
this one arrives at a length for AJ, the presumed circumference of a disk of 
radius r, of precisely as predicted by Einstein and many others

\begin{equation}
\label{eq4}
{\rm A}{\rm J} = {\frac{{2\pi r}}{{\sqrt {1 - \omega ^{2}r^{2} / c^{2}} 
}}}{\rm .}
\end{equation}

But consider that since point p moves along a timelike path as seen from K 
(see dotted line in Fig. 3), a time difference between events A and J must 
therefore exist as measured by a clock attached to point p. As a result, one 
end of a continuous tape measure riding with the rim of the disk would not 
meet back up with its other end at the same point in time. But any 
meaningful measurement of the circumference simply must have the same 
starting and ending event, and therefore must be a \textit{closed path} in spacetime.

We have therefore shown that the tangent frames analysis approach leads to a 
discontinuity in time, a seemingly impossible physical situation. Even 
further, the spatial location of that discontinuity is completely arbitrary. 
It depends on where we choose our initial starting point p. This is a very 
serious dilemma for the traditional interpretation.

We conclude that simultaneity can not be defined in a consistent manner 
using local inertial clocks over any closed path where different parts of 
the path have different relative velocities. Hence, we are unable to measure 
the circumference of the rim with local inertial rods where the endpoints of 
all the rods are simultaneous as measured by local inertial clocks. 
Therefore, inertial frames tangent to the rim can not be used to measure the 
disk circumference, and conclusions made from so doing will not be valid.

This apparent violation of the heretofore seemingly sacrosanct "surrogate 
rods postulate" is addressed in Sec. 5.2.

\subsection*{2.4 Related Problems}

Ehrenfest \cite{Ehrenfest:1909,Sama:1972} saw a paradox in the 
presumed circumferential Lorentz contraction effect which Einstein 
\cite{Ref:7} and Gr{\o}n \cite{Relativistic:1975} attempted to resolve 
by claiming that the disk circumference tries to contract in Lorentz 
fashion, but can't, and so undergoes internal tensile stress. Other 
mechanically induced stresses aside, the disk presumably cannot be spun up 
to relativistic speeds without developing such stresses and, at high enough 
speeds, rupturing. 

But one must then also argue that the time discontinuity of Fig. 3 is 
resisted in some way by a "tension" in the time component around the 
circumference. If the rods must be extended in order to meet up in space, 
then surely the endpoints of rods must be adjusted in time in order to meet 
up as well. Tensile stress may be a well known physical phenomenon in a 
material body in space, but there is certainly no such phenomenon associated 
with time.

A related problem is pointed out by several authors (see, for example Weber 
[13]). If light rays sent out around the circumference are used according to 
the standard Einstein synchronization procedure, one finds the clock at p at 
360\r{}  to be out of synchronization with itself at 0\r{} . This leads to 
the restriction that one can only consider open paths on the disk surface. 
But then one must ask what prevents a physical disk based observer from 
traveling around one complete circumference? And what prevents her from 
laying down a continuous tape measure as she does so? And finally, how good 
a representation of the physical world is a model in which a clock can not 
be synchronized with itself?

\section*{3 NEW THEORY OF ROTATING FRAMES}

In this section we re-derive key aspects of relativity theory for the 
rotating reference frame using two new postulates based on the Sagnac and 
other experiments. We follow logic similar to that employed by Einstein to 
derive special relativity for translational motion, but start from a 
different, but equally empirically justifiable, basis.

In Sec. 4 transformation techniques of differential geometry are utilized to 
rigorously derive all relevant characteristics of the rotating frame, 
including the exact form of Eq. (\ref{eq2}), our first postulate below. The present 
Sec. 3, on the other hand, provides a physically meaningful, and simpler, 
derivation of certain of those characteristics.

\subsection*{3.1 New Postulates}

We postulate the following:

1. The speed of light is not invariant between the ground and the rotating 
frame, and in the rotating frame is found to first order by the velocity 
addition law of Eq. (\ref{eq2})

\[
\vert v_{{\rm l}{\rm i}{\rm g}{\rm h}{\rm t}} {\rm \vert} {\rm} {\rm}  \cong 
{\rm} {\rm} c{\rm} {\rm} \pm {\rm} {\rm} \omega r
\]

2. Observers can discern which frame is non-rotating (the "preferred 
frame").

\subsection*{3.2 Different Results}

\subsubsection*{3.2.1 Simultaneity.}

Fig. 4 depicts a means for defining simultaneity at any radius r on the 
disk. Light rays can be imagined as emitted simultaneously from the 
centerpoint of the disk, striking mirrors located at $r,$ and being reflected 
back to the centerpoint. They all arrive back at the center at the same 
instant in time as measured by a clock located there, and one concludes that 
the events occurring when the light struck the mirrors are all simultaneous. 
These events are also simultaneous to observers in K.

\begin{figure}
\centering
\includegraphics*[bbllx=0.26in,bblly=0.13in,bburx=5.22in,bbury=2.77in,scale=1.00]{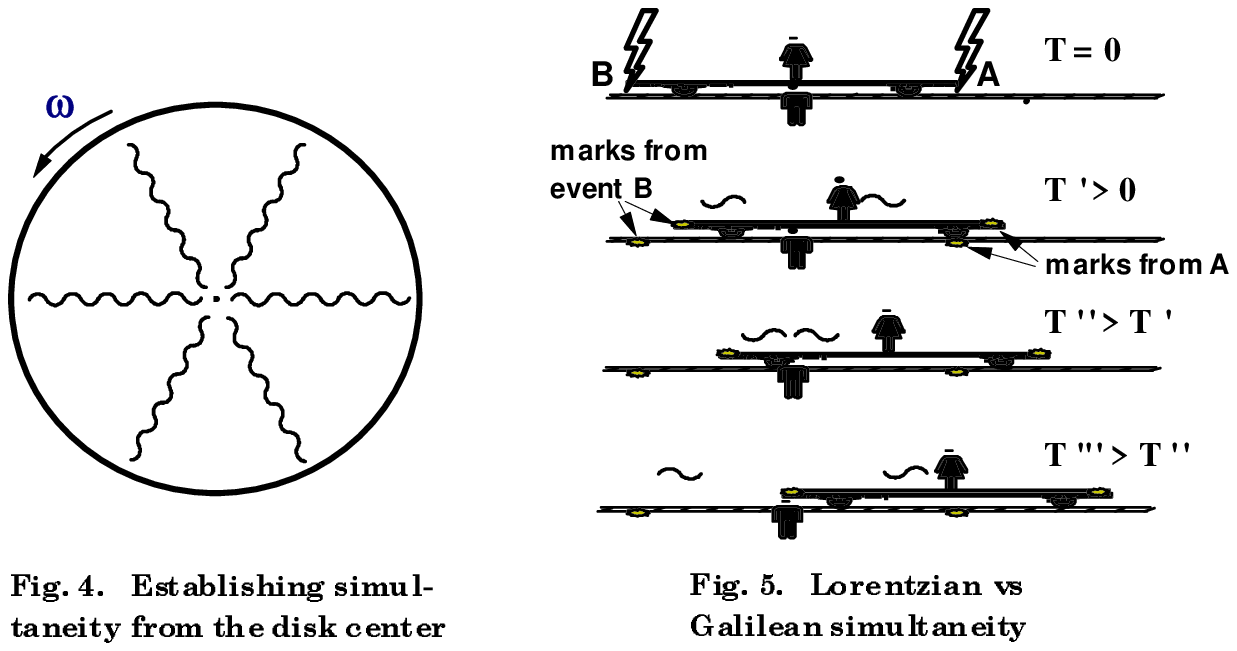}
\end{figure}

\bigskip

The question then arises as to whether a non-inertial observer riding on the 
rim itself would agree that those same events are simultaneous. Standard 
relativity theory predicts she would not, since she and the K observer have 
relative velocity difference.

\bigskip

We answer this question by re-considering Einstein's famous \textit{gedanken} experiment of 
the passing train shown in Fig. 5. Lightning strikes both ends of the car 
and leaves marks on both ends plus the ground. These events are A and B. 
Given the postulate that light has the same velocity as seen from the train 
or the ground, and given that both observers can later measure the distance 
to the brown marks left by lightning events and determine that each is 1/2 
way between their respective marks, the train observer concludes that A 
occurred before B because she saw the A flash of light first. This she 
concludes because she knows the speed of light from both directions is the 
same for her. The ground observer sees each flash at the same instant and 
concludes the two events were simultaneous since the speed of light is also 
the same for him in both directions. This is case 1, the standard special 
relativity result.

For case 2, suppose instead that nature works in Galilean fashion and the 
light from A travels faster than the light from B as seen from the train 
frame. ($\vert $V$_{{\rm A}}\vert $ = c + $v$ and $\vert $V$_{{\rm B}}\vert 
$ = c - $v$ where $v$ is the absolute value of the relative velocity between 
frames.) The train observer still sees the A flash first, and still later 
measures the distance to the marks and knows she is 1/2 way between them. 
But now she also knows that the light from A travels faster, so she would 
expect to see it first. The math is trivial. She concludes that A and B were 
indeed simultaneous, as does the ground observer.

As shown by the Sagnac experiment the speed of light on the circumference of 
the disk behaves as in the second case above. (Assume for the present that 
Eq. (\ref{eq2}) is an exact equality. We will resolve the first order approximation 
issue in Sec. 4.) The observer on the disk knows she is rotating, knows she 
has tangential velocity relative to the inertial frame K, and knows from Eq. 
(\ref{eq2}) the formula for calculating the velocity of light as seen by her (it is 
direct addition as for the train case 2 above.) She therefore concludes that 
two spatially proximate events on the circumference which are simultaneous 
in the ground frame are also simultaneous to her even though she sees one of 
them occur first.

Hence, whether measured from the center of the disk, or locally at any other 
point on the disk, simultaneity in the disk frame k is identical with that 
of K. (Selleri [18] agrees with this conclusion, although he takes a 
different route to get there.) So unlike systems with relative rectilinear 
velocities where there is no common agreement in simultaneity, systems with 
relative rotational velocities all do agree on simultaneity.

\subsubsection*{3.2.2 No Lorentz Contraction.}

The Lorentz contraction is a direct result of non-agreement in simultaneity 
between frames. If there is agreement in simultaneity, there is no Lorentz 
contraction. To show this we need one additional, presumably inviolable, 
postulate. That is,

\bigskip

3. The proper spacetime length of any path is invariant under any 
transformation, i.e., it is the same for all observers.

\bigskip

Hence, for two frames in relative motion (notation should be obvious)

\begin{equation}
\label{eq5}
{\rm} {\rm} {\rm (}\Delta s)^{2} = {\rm}  - c^{2}(\Delta t)^{2} + (\Delta 
l)^{2} = {\rm}  - c^{2}(\Delta t')^{2} + (\Delta l')^{2}
\end{equation}

For a rod at rest in the primed frame, an observer in the unprimed frame 
sees that rod such that its endpoints are \textit{events} which for him occur 
simultaneously, i.e., $\Delta t$ = 0. But in the primed system those events 
are not, according to standard relativity theory, simultaneous and $\Delta 
{t}' \ne 0$. This means $\Delta l \ne \Delta {l}'$, and results in Lorentz 
contraction \cite{There:1}.

If, however, the same two events could also appear simultaneous in the 
primed system, then $\Delta t'$ = 0, and $\Delta l $must equal $\Delta l'$. This 
is, of course, not possible for two frames in relative \textit{translational} motion, but, as we 
have shown, it is possible between two frames with different \textit{rotational} motion.

Hence, if $\Delta l' $is the length of a (short) standard measuring rod 
attached to the non-rotating K frame aligned tangentially to the disk rim, 
and $\Delta l$ is the length of a similar rod attached tangentially to the 
rim, then neither rod looks shortened to observers in either frame. There 
is, therefore, no Lorentz contraction for rotating systems.

The reader should note carefully the distinction here between with the 
contention of Gr{\o}n [10] and others \cite{Arzeli:1966} that rods fixed 
to the disk will not contract since tension in the disk prevents them from 
so doing. In contradistinction, we show that there is simply no kinematic 
imperative for the rods to try to contract. No tension arises in the disk as 
it is spun up, and no relativistically induced rupturing occurs.

\subsubsection*{3.2.3 Time Dilation.}

Although frames K and k agree on simultaneity, it can be shown that standard 
clocks in each run at different rates. (Note that two clocks running at 
different rates can nonetheless both agree on simultaneity, i.e., that no 
time elapsed off either one between two events.)

The time dilation effect can be demonstrated with the aid of the spacetime 
diagram of Fig. 6, which shows the helical path of a clock fixed on the disk 
as seen from frame K and that of a clock fixed in K as seen from K. The 
moving clock travels the path of 3D point p of Fig. 3 extended for one full 
rotation. The path of that clock is a non-geodesic, while the path of its 
"twin" fixed in K is a geodesic, a straight line. The proper time passed for 
each clock is simply its path length (divided by \textit{ic}), and this path length can 
be measured in any frame we choose since it is frame invariant. We choose 
frame K since it is the simplest.

In frame K, the proper \textit{spatial} distance traversed by the disk fixed clock is 
$\Delta \sigma $ = 2$\pi R = $($\omega \Delta T$)$R$ where the time interval for 
one rotation is $\Delta T. $Hence, the proper spacetime path length of the 
moving clock is

\begin{equation}
\label{eq6}
{\rm (}\Delta s)^{2}{\rm}  = - \,c^{2}(\Delta \tau )^{2} = \; - c^{2}(\Delta 
T)^{2} + (\Delta \sigma )^{2}{\rm}  = - \,c^{2}(\Delta T)^{2} + \omega 
^{2}R^{2}(\Delta T)^{2}
\end{equation}

Hence,

\begin{equation}
\label{eq7}
\Delta \tau {\rm} {\rm}  = {\rm} \Delta t{\rm} {\rm}  = {\rm} (1{\rm}  - 
{\rm} r^{2}\omega ^{2} / c^{2})^{1 / 2}\Delta T = {\rm} (1{\rm}  - {\rm 
}v^{2} / c^{2})^{1 / 2}\Delta T
\end{equation}

\begin{figure}
\centering
\includegraphics*[bbllx=0.26in,bblly=0.13in,bburx=3.34in,bbury=2.59in,scale=1.00]{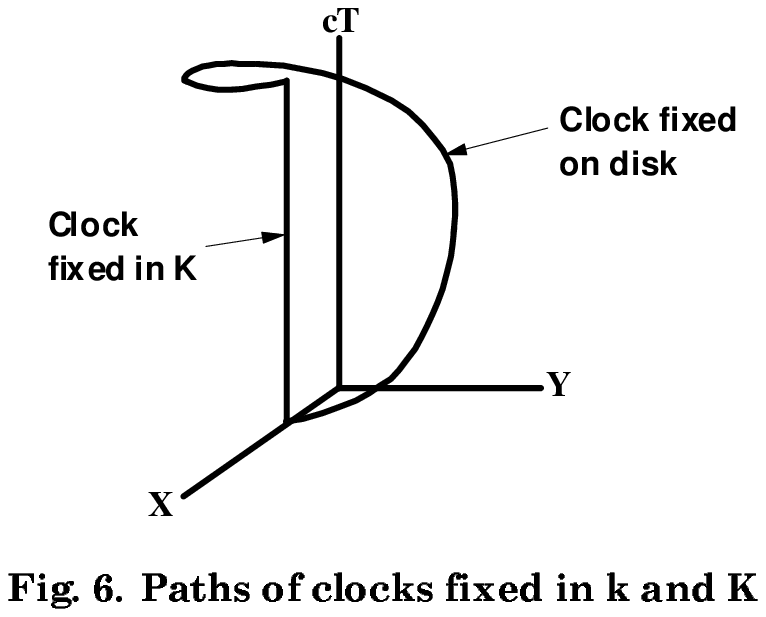}
\end{figure}

\bigskip

\noindent
and the clock fixed in k on the disk rim runs slower than the K clock on the 
ground. Also, clocks run more slowly at greater radii, so it is not possible 
to synchronize standard clocks at different radii.

\bigskip

It is noteworthy that similar time dilation effects occur in translationally 
accelerating systems, yet it is readily shown [22] that such systems are 
nevertheless Riemann flat. (Acceleration does not cause spacetime curvature, 
gravity does.) Hence, time dilation in and of itself is not a sufficient 
condition for curvature.

Note also that, analogous with the rectilinearly accelerating system, it is 
not possible to synchronize standard clocks located at different radii in k 
since such clocks beat at different rates.

The analysis of a clock fixed on the disk at a certain radius r is similar 
to that of the traveling twin in the classic "twin paradox". Both twins live 
in Minkowski spaces, but the traveling twin follows a non-geodesic in 
spacetime (it must decelerate/accelerate to return to earth) and hence has a 
shorter elapsed time than the geodesic following sibling. This result, as in 
the rotating disk case, is independent of the reference frame of the 
observer since proper pathlength is invariant under transformation.

\section*{4 TRANSFORMATION THEORY}
\label{sec:mylabel1}

Einstein used his two postulates to derive the Lorentz transformation, from 
which all relevant relativistic characteristics may be found. If, 
conversely, he had known the Lorentz transformation first, he could have 
then derived his two postulates. In the present sec., we start with a 
reasonable guess at the correct transformation between rotating frames, 
analogous to the Lorentz transformation between translationally moving 
frames, and not only derive our original postulates, but predict other 
phenomena as well. As will be shown, these other phenomena are self 
consistent, do not lead to the difficulties delineated in Sec. 2, and agree 
with all known experiments.

\subsection*{4.1 Rotating Frame Metric and Transformations}

Strauss [9], Franklin \cite{Franklin:1922}, Trocheries 
\cite{Trocheries:1949}, and Takeno \cite{Takeno:1952} have attempted 
to impose transformations between inertial and rotating frames which make an 
\textit{a priori} assumption that the Lorentz contraction is operative and varies with the 
radius $r$ of the disk (i.e., varies with the tangential velocity in the 
traditional special relativistic manner). These transformations appear to 
put the cart before the horse, i.e., they \textit{start} with the Lorentz contraction 
built in.

An alternative, and more reasonable transformation (see Eqs. (8a-d) below) 
found in many sources 
[10,11,12,13,16,26,\cite{Pellegrini:1995},\cite{Ridgely:1998}] (although 
with different interpretations and results than the present paper) makes no 
such assumption. It simply makes kinematic connections between the 
cylindrical rotating and non-rotating coordinate systems which are 
straightforward and seem most logical. If the transformation is correct, 
appropriate effects derivable from it should agree with experiment, and 
predicted results should be self consistent.

This coordinate transformation, where upper case coordinates represent the 
inertial frame K, lower case denote the rotating frame k, and the axis of 
rotation is coincident with both the $Z $and $z$ axes, is

\begin{equation}
\label{eq8}
\begin{array}{l}
 cT = ct\quad \quad \quad \quad \quad {\rm (}{\rm 8}{\rm a}{\rm )} \\ 
 R = r\quad \quad \quad \quad \;\;\;\quad {\rm (}{\rm 8}{\rm b}{\rm )} \\ 
 \Phi = \phi + \omega t\quad \quad \quad \;\;{\rm (}{\rm 8}{\rm c}{\rm )} \\ 
 Z = z\quad \quad \quad \quad \quad \;\;\;{\rm (}{\rm 8}{\rm d}{\rm )} \\ 
 \end{array}
\end{equation}

 $\omega $ is the angular velocity of the disk, and $t,$ the coordinate time for the 
rotating system, is the proper time of a standard clock located at the 
origin of the rotating coordinate frame, i.e., it is equivalent to any 
standard clock at rest in K. Note that $t$ is only a coordinate. It is merely a 
label and cannot be expected to equal proper time at any given point on the 
disk (except, of course, at $r $= 0).

Assumptions upon which transformation (\ref{eq8}) is based are: (i) The radial 
distance $r$ measured in k can not be contracted as seen from K since velocity 
is always perpendicular to $R$, hence $R$ = $r$. (ii) Radii in k (i.e., lines of 
constant $\phi $ and constant $z$) each are straight lines as seen from either 
k or K, move with rotational velocity $\omega $, and are independent of $r$. 
(iii) The rotation has no effect on measurements in the direction of the 
axis of rotation, i.e., the $Z$ direction, since, like the radial distance, $Z = z$ is 
perpendicular to velocity. Assumptions (i) and (iii) are apparently 
universally accepted by others. Assumption (ii) leads to Eq. (8c) and, as 
mentioned, has been considered by others.

The transformation (\ref{eq8}) seems Galilean in nature, rather than relativistic, 
and if it is valid (as most researchers today feel that it is), we should 
not be surprised to find the disk exhibiting at least some Galilean 
characteristics.

To deduce the metric for the rotating system we begin with the line element 
for the standard cylindrical coordinate system of the Minkowski space K

\begin{equation}
\label{eq9}
ds^{2}{\rm}  = {\rm}  - {\rm} c^{2}dT^{2} + {\rm} dR^{2}{\rm}  + {\rm 
}R^{2}d\Phi ^{2}{\rm}  + {\rm} dZ^{2}{\rm .}
\end{equation}

Finding \textit{dT, dR, d}$\Phi ,$ and \textit{dZ} from Eqs. (\ref{eq8}), and inserting into Eq. (\ref{eq9}), one obtains 
the metric of the coordinate grid in k. (Note this step incorporates 
postulate 3 of Sec. 3.2.2, i.e., \textit{ds} is invariant.)

\begin{equation}
\label{eq10}
\begin{array}{c}
 ds^{2}\;\; = \; - c^{2}(1 - {\textstyle{{r^{2}\omega ^{2}} \over 
{c^{2}}}})dt^{2}\; + \;dr^{2}\; + \;r^{2}d\phi ^{2}\; + \;2r^{2}\omega d\phi 
dt\; + \;dz^{2} \\ 
 \\ 
 = \;g_{\alpha \beta}  dx^{\alpha} dx^{\beta} \quad , \\ 
 \end{array}
\end{equation}

\noindent
where the covariant form of the metric $g_{{\rm \alpha} {\rm \beta} }$ and 
its inverse, the contravariant matrix $g^{{\rm \alpha} {\rm \beta} }$, 
readily found via the standard cofactor method, are

\begin{equation}
\label{eq11}
g_{\alpha \beta}  = {\left[ {{\begin{array}{*{20}c}
 { - (1 - {\textstyle{{r^{2}\omega ^{2}} \over {c^{2}}}})} \hfill & {0} 
\hfill & {{\textstyle{{r^{2}\omega}  \over {c}}}} \hfill & {0} \hfill \\
 {0} \hfill & {1} \hfill & {0} \hfill & {0} \hfill \\
 {{\textstyle{{r^{2}\omega}  \over {c}}}} \hfill & {0} \hfill & {r^{2}} 
\hfill & {0} \hfill \\
 {0} \hfill & {0} \hfill & {0} \hfill & {1} \hfill \\
\end{array}} } \right]}\quad g^{\alpha \beta}  = {\left[ 
{{\begin{array}{*{20}c}
 { - 1} \hfill & {0} \hfill & {{\textstyle{{\omega}  \over {c}}}} \hfill & 
{0} \hfill \\
 {0} \hfill & {1} \hfill & {0} \hfill & {0} \hfill \\
 {{\textstyle{{\omega}  \over {c}}}} \hfill & {0} \hfill & {(1 - 
{\textstyle{{r^{2}\omega ^{2}} \over {c^{2}}}}) / r^{2}} \hfill & {0} \hfill 
\\
 {0} \hfill & {0} \hfill & {0} \hfill & {1} \hfill \\
\end{array}} } \right]}{\rm .}
\end{equation}

For future reference, the comparable matrices in K are

\begin{equation}
\label{eq12}
G_{AB} = {\left[ {{\begin{array}{*{20}c}
 { - 1} \hfill & {0} \hfill & {0} \hfill & {0} \hfill \\
 {0} \hfill & {1} \hfill & {0} \hfill & {0} \hfill \\
 {0} \hfill & {0} \hfill & {R^{2}} \hfill & {0} \hfill \\
 {0} \hfill & {0} \hfill & {0} \hfill & {1} \hfill \\
\end{array}} } \right]}\quad \quad \quad G^{AB} = {\left[ 
{{\begin{array}{*{20}c}
 { - 1} \hfill & {0} \hfill & {0} \hfill & {0} \hfill \\
 {0} \hfill & {1} \hfill & {0} \hfill & {0} \hfill \\
 {0} \hfill & {0} \hfill & {{\textstyle{{1} \over {R^{2}}}}} \hfill & {0} 
\hfill \\
 {0} \hfill & {0} \hfill & {0} \hfill & {1} \hfill \\
\end{array}} } \right]}
\end{equation}

\noindent
where sub and superscripts $A$ and $B$ as used here are upper case Greek letters 
for alpha and beta.

Note from Eqs. (\ref{eq10}) and Eqs. ((\ref{eq11}) that the rotating disk system is not 
orthogonal (the metric is not diagonal).

Taking the differentials in Eqs. (\ref{eq8}), one can readily derive the matrix 
$\Lambda ^{{\rm \alpha} }_{{\rm {\rm B}}}$ which transforms 
contravariant components of vectors and tensors from K to k, and its inverse 
$\Lambda ^{{\rm {\rm A}}}_{{\rm \beta} }$ which transforms contravariant 
vectors and tensors from k to K. These transformations between the two 
cylindrical coordinate systems are:

\begin{equation}
\label{eq13}
\Lambda ^{\alpha} _{B} = {\left[ {{\begin{array}{*{20}c}
 {1} \hfill & {0} \hfill & {0} \hfill & {0} \hfill \\
 {0} \hfill & {1} \hfill & {0} \hfill & {0} \hfill \\
 { - {\textstyle{{\omega}  \over {c}}}} \hfill & {0} \hfill & {1} \hfill & 
{0} \hfill \\
 {0} \hfill & {0} \hfill & {0} \hfill & {1} \hfill \\
\end{array}} } \right]}\;,\quad \quad \Lambda ^{A}_{\beta}  = {\left[ 
{{\begin{array}{*{20}c}
 {1} \hfill & {0} \hfill & {0} \hfill & {0} \hfill \\
 {0} \hfill & {1} \hfill & {0} \hfill & {0} \hfill \\
 {{\textstyle{{\omega}  \over {c}}}} \hfill & {0} \hfill & {1} \hfill & {0} 
\hfill \\
 {0} \hfill & {0} \hfill & {0} \hfill & {1} \hfill \\
\end{array}} } \right]}\;.
\end{equation}

With the above metrics and transformations forming the basis of the new 
theory, we can proceed to derive the effects we would expect to see in the 
physical world.

\subsection*{4.2 Galilean Characteristics}
\label{subsec:mylabel3}

\subsubsection*{4.2.1 Invariance of simultaneity.}

From Eq. (8a) [or equivalently by comparing Eq. (\ref{eq9}) and Eq. (\ref{eq10})], if 
$\Delta T = $0 in K for two events, then $\Delta t$ = 0 in k for the same two 
events. In other words the two events are simultaneous as seen from either 
system, in agreement with our earlier thought experiment based on physical 
reasoning. Note that the transformation between time coordinates of the 
present theory is much different than that of the Lorentz transformation. In 
the latter the coordinate time difference is dependent upon the locations of 
the two events; in the former, it is not.

\subsubsection*{4.2.2 No Lorentz contraction.}

Note since Eq. (\ref{eq9}) equals Eq. (\ref{eq10}), the circumference of the disk for any 
radius $r = R$, at fixed time $t $and constant $z $(i.e., \textit{dt = dT = dr = dR = dz = dZ =} 0)$,$ is 2$\pi r $($\Delta \Phi $ = 
$\Delta \phi $ = 2$\pi $), implying that the rotating disk is indeed a 
flat space, and corroborating the physical reasoning of Sec. 3.2.2. (Some 
authors, most notably Gr{\o}n [10,26], contend that the non-time orthogonal 
nature of the rotating coordinate system negate this conclusion. We resolve 
this matter in Sec. 5 below and the Appendix.)

Note further that Lorentz contraction arises directly from the Lorentz 
transformation, yet Eqs. (\ref{eq8}), the transformation now accepted as correct by 
virtually everyone in the field, is not the Lorentz transformation. There 
is, therefore, absolutely no reason (other than tradition) to tacitly assume 
that it must somehow give rise to Lorentz contraction.

\subsubsection*{4.2.3 Angular velocity addition.}

Consider three co-axial reference frames, one of which is not rotating and 
designated by K, the second of which has rotational velocity $\omega 
_{{\rm 2}}$ and designated by k$_{{\rm 2}}$, and the third of which has 
velocity $\omega _{{\rm 3}}$ = 2$\omega _{{\rm 2}}$ and is designated by 
k$_{{\rm 3}}$. $\omega _{{\rm 2}}$ and $\omega _{{\rm 3}}$ are measured 
relative to K. Note that an observer in k$_{{\rm 2}}$ sees the k$_{{\rm 3}}$ 
system rotate once relative to him, for each time interval that he rotates 
once relative to K. Hence $\omega _{{\rm 3}{\rm /} {\rm 2}}$, the angular 
velocity of k$_{{\rm 3}}$ relative to k$_{{\rm 2}}$, has the same magnitude 
as $\omega _{{\rm 2}{\rm} }$, and therefore

\begin{equation}
\label{eq14}
\omega _{3} {\rm}  = {\rm} {\rm} \omega _{2} {\rm}  + {\rm} \omega _{3 / 2} 
{\rm} 
\end{equation}

Relationship Eq. (\ref{eq14}) obviously holds in general, and demonstrates that 
rotational velocities for co-axial systems add directly, in Galilean 
fashion, frame to frame and not relativistically as do translational 
velocities. This not only lends further credence to the Galilean type 
transformation (\ref{eq8}) employed herein, but also implies that there is no upper 
limit on angular velocity comparable to the luminal limitation on 
rectilinear velocities \cite{will:1}.

\subsubsection*{4.2.4 Translational velocity addition.}
\label{subsubsec:mylabel2}

Assume $V^{{\rm I}}$ are the components of the three velocity of an object as 
seen in the K cylindrical coordinate system, and $U^{{\rm A}}$ are the 
components of the four-velocity. For the same object, an observer in k 
measures $v^{i}$\textit{} as components of the three velocity and $u^{{\rm \alpha} }$ 
for the four-velocity. That is,

\begin{equation}
\label{eq15}
V^{I} = {\frac{{dX^{I}}}{{dT}}}\;,\quad \quad U^{A} = 
{\frac{{dX^{A}}}{{d\tau} }} = {\textstyle{{1} \over {\sqrt {1 - v^{2} / 
c^{2}}} }}{\left[ {{\begin{array}{*{20}c}
 {c} \hfill \\
 {{\textstyle{{dR} \over {dT}}}_{_{}}  = V^{R}} \hfill \\
 {{\textstyle{{d\Phi}  \over {dT}}}_{_{}}  = V^{\Phi} } \hfill \\
 {{\textstyle{{dZ} \over {dT}}}_{_{}}  = V^{Z}} \hfill \\
\end{array}} } \right]} \quad {\rm .}
\end{equation}

To find the three velocity addition law, we use the same procedure employed 
in special relativity to derive the relativistic velocity addition law. We 
begin by first transforming the four-vector \textit{dX}$^{{\rm I}}$ to its counterpart 
in k, \textit{dx}$^{{\rm \alpha} }$, using the appropriate transformation matrix from 
Eqs. (\ref{eq13}).

\begin{equation}
\label{eq16}
{\left[ {{\begin{array}{*{20}c}
 {cdt} \hfill \\
 {dr} \hfill \\
 {d\phi}  \hfill \\
 {dz} \hfill \\
\end{array}} } \right]} = dx^{\alpha}  = \Lambda ^{\alpha} _{B} dX^{B} = 
{\left[ {{\begin{array}{*{20}c}
 {1} \hfill & {0} \hfill & {0} \hfill & {0} \hfill \\
 {0} \hfill & {1} \hfill & {0} \hfill & {0} \hfill \\
 { - {\textstyle{{\omega}  \over {c}}}} \hfill & {0} \hfill & {1} \hfill & 
{0} \hfill \\
 {0} \hfill & {0} \hfill & {0} \hfill & {1} \hfill \\
\end{array}} } \right]}{\left[ {{\begin{array}{*{20}c}
 {cdT} \hfill \\
 {dR} \hfill \\
 {d\Phi}  \hfill \\
 {dZ} \hfill \\
\end{array}} } \right]} = {\left[ {{\begin{array}{*{20}c}
 {cdT} \hfill \\
 {dR} \hfill \\
 { - \omega dT + d\Phi}  \hfill \\
 {dZ} \hfill \\
\end{array}} } \right]}
\end{equation}

Three velocities in k are then found simply by dividing the spatial 
components of Eq. (\ref{eq16}) by \textit{dt,} and noting that \textit{dt = dT}, i.e.,

\begin{equation}
\label{eq17}
v^{i} = {\frac{{dx^{i}}}{{dt}}} = {\left[ {{\begin{array}{*{20}c}
 {{\textstyle{{dR} \over {dt}}}} \hfill \\
 { - \omega {\textstyle{{dT} \over {dt}}} + {\textstyle{{d\Phi}  \over 
{dt}}}} \hfill \\
 {{\textstyle{{dZ} \over {dt}}}} \hfill \\
\end{array}} } \right]} = {\left[ {{\begin{array}{*{20}c}
 {V^{R}} \hfill \\
 { - \omega + V^{\Phi} } \hfill \\
 {V^{Z}} \hfill \\
\end{array}} } \right]}
\end{equation}

For an object with purely tangential velocity of magnitude $V^{{\rm T}{\rm 
a}{\rm n}{\rm g}}$ equal to \textit{RV}$^{{\rm \Phi} }$ one finds from Eq. (\ref{eq17}) that

\begin{equation}
\label{eq18}
{\rm} v^{{\rm t}{\rm a}{\rm n}{\rm g}}{\rm}  = {\rm} {\rm} rv^{\phi} {\rm}  
= {\rm}  - {\rm} \omega R{\rm} {\rm}  + {\rm} RV^{\Phi} {\rm} {\rm =} {\rm 
}{\rm} {\rm -} {\rm} \omega R{\rm}  + {\rm} V^{{\rm t}{\rm a}{\rm n}{\rm 
g}}
\end{equation}

\noindent
a very Galilean-looking transformation.

It must be noted once again, however, that time derivatives above are with 
respect to coordinate time $t, $and for a disk fixed observer at any location 
other than $r$ = 0, time dilation effects must be taken into account to reflect 
the actual velocities such an observer would measure with physical 
instruments. In practice this would mean dividing Eq. (\ref{eq18}) by the factor 
$\sqrt {1 - \omega ^{2}r^{2} / c^{2}} $, i.e., by the factor local time 
differs from the coordinate time used in Eq. (\ref{eq17}). Note that this does not 
change the directly additive quality of Eq. (\ref{eq18}).

\subsubsection*{4.2.5 Lack of invariance of the speed of light.}
\label{subsubsec:mylabel3}

Consider Eq. (\ref{eq18}) where $V^{{\rm T}{\rm a}{\rm n}{\rm g}}$ represents the 
speed $c$ of a light ray which could be propagating in the positive or negative 
$\Phi $ direction. Then

\begin{equation}
\label{eq19}
v^{{\rm l}{\rm i}{\rm g}{\rm h}{\rm t}{\rm ,}{\rm t}{\rm a}{\rm n}{\rm g}} = 
\;\; - \omega R\;\;\pm \;\;c
\end{equation}

This result is in remarkable agreement with the Sagnac experiment, and 
provides strong support for the validity of transformation (\ref{eq8}).

Since velocities in Eq. (\ref{eq19}) are coordinate velocities, we must divide both 
sides of the equation by the time dilation factor $\sqrt {1 - \omega 
^{2}r^{2} / c^{2}} $ to represent the physical velocities a disk observer 
would actually measure using local standard clocks. By doing this we obtain 
the exact relationship for which the Sagnac result Eq. (\ref{eq2}) was only a first 
order approximation. Hence, as we assumed in Sec. 3.2.1, the exactly equal 
sign in Eq. (\ref{eq2}) is correct if the velocities are taken as those which would 
actually be measured by an observer fixed to the disk; see Sec. 5.1 and Eq. 
(\ref{eq33}).

By utilizing these velocities, rather than $c$, for light rays employed to 
synchronize clocks at a given radius, one then finds a clock at 360$^{o}$ is 
synchronized with itself at 0$^{o}$. More generally, closed path 
integrations are fully allowable, and thereby consistent with what one would 
expect physically.

Note that we have derived our rotating frame postulates from transformation 
(\ref{eq8}). Eq. (\ref{eq19}) is the first postulate. By taking $\omega $ = 0 in Eq. (\ref{eq19}), 
we get our second. That is, the preferred frame is the one with isotropic 
light speed $c$, i.e., it is the inertial one.

In Sec. 5.1 we re-derive these results even more rigorously, and reconcile 
them with general relativity principle 1 in Sec. 1.2.

\subsection*{4.3 Lorentzian Characteristics}
\label{subsec:mylabel4}

A plethora of cyclotron experiments demonstrates that rotating systems do 
indeed possess certain relativistic characteristics, such as time dilation 
(longer decay times) and mass-energy increase with speed. If transformation 
(\ref{eq8}) is the correct one, then these effects must be predicted by it. The 
ensuing derivations do indeed confirm that transformation (\ref{eq8}) is consistent 
with these experiments.

\subsubsection*{4.3.1 Time dilation.}

From the metric of Eq. (\ref{eq10}) with r, $\phi $, and z constant and \textit{ds}$^{{\rm 2}}$ 
= $- c^{{\rm 2}}d\tau ^{{\rm 2}}$, the proper time at any radius $r$ is

\begin{equation}
\label{eq20}
d\tau \;\; = \quad \sqrt {1\; - \;{\textstyle{{r^{2}\omega ^{2}} \over 
{c^{2}}}}} \;dt\quad = \quad \sqrt {1\; - \;{\textstyle{{v^{2}} \over 
{c^{2}}}}} \;dT
\end{equation}

\noindent
which corroborates the result Eq. (\ref{eq7}) of Sec. 3.2.3. (Since speed is 
constant, finite "$\Delta $" differences in Eq. (\ref{eq7}) can be taken over to 
differentials "$d$".)

Note that the time dilation effect arises naturally from the simple and 
readily justifiable coordinate transformation (\ref{eq8}), and was not "built in" 
from the start by assuming that it holds \textit{a priori.} Further, time dilation does occur 
in the rotating frame in accordance with the standard relation of special 
relativity. However, unlike special relativity this effect is not symmetric 
between frames k and K. Observers in both systems agree that the rotating 
disk clocks run slower.

\subsubsection*{4.3.2 Path lengths of light and particles.}
\label{subsubsec:mylabel4}

The pathlength of any object traveling in spacetime is invariant between 
frames in accordance with our "new" postulate 3, which is, of course, not 
really new but a fundamental principle of differential geometry. The path 
length of light, in particular, remains null as viewed from the rotating 
frame since \textit{ds} = 0 in K, and hence \textit{ds} must also = 0 in k.

\subsubsection*{4.3.3 Four-vectors.}

The four-velocity and the four-momentum transform readily between the 
rotating and non-rotating systems also in accordance with basic principles 
of differential geometry/general relativity.

However, when making general transformations of four-vectors, one should 
keep two things in mind which are usually irrelevant for Minkowski metrics 
in Minkowski space, but are quite relevant for other metrics such as that of 
the rotating frame. Both of these relate to physical interpretation of the 
components of four-vectors (i.e., the quantities one would actually measure 
with instruments.)

The first of these concerns lies with the covariant or contravariant nature 
of the components. Since coordinate differences (e.g., \textit{dx}$^{{\rm \alpha} }$) 
are expressed as contravariant quantities, and since four-velocity is simply 
the derivative of these coordinate differences with respect to the invariant 
scalar quantity $\tau $ (proper time), four-velocities only represent 
(proper) time derivatives of the coordinate values if they are expressed in 
contravariant form. In general, lowering the index of $u^{{\rm \alpha} }$ via 
the metric $g_{{\rm \alpha} {\rm \beta} }$ gives components $u_{{\rm \alpha 
}}$ which are \textit{not} the time derivatives of the coordinate values. This is true 
because $g_{{\rm \alpha} {\rm \beta} }$ is not the identity matrix. Note that 
in inertial frames $g_{{\rm \alpha} {\rm \beta} }$ =\textit{} $\eta _{{\rm \alpha 
}{\rm \beta} }$ (see Eq. (\ref{eq1})) which is, apart from the sign of the 
$g_{{\rm 0}{\rm 0}}$ component, an identity matrix. In a coordinate frame 
with such a Minkowski metric the covariant form of the four-velocity is 
identical to the contravariant form except for the sign of the timelike 
component. In other coordinate frames, however, the difference is much more 
significant, and care must be taken to work with the contravariant form of 
the four-velocity.

Four-momentum, on the other hand, must be treated in terms of its covariant 
components. This is because said four-momentum is the canonical conjugate of 
the four-velocity. In brief, if the Lagrangian of a given system is

\begin{equation}
\label{eq21}
L\;\; = \;\;L(x^{\alpha} ,\dot {x}^{\alpha} ,\tau )
\end{equation}

\noindent
where dots over quantities represent derivatives with respect to $\tau $, 
then the conjugate momentum is

\begin{equation}
\label{eq22}
p_{\alpha}  \; = \;{\frac{{\partial L}}{{\partial \dot {x}^{\alpha} }}}
\end{equation}

Hence it is imperative that one use the covariant components of the 
four-momentum. Contravariant components, for all but a Minkowski metric, 
will not represent physical quantities such as energy, three momentum, etc.

Getting the correct contravariant or covariant components is not quite 
enough, however, in order to compare theoretical results with measured 
quantities. If a given basis vector does not have unit length, the magnitude 
of the corresponding component will not equal the physical quantity 
measured. For example, a vector with a single non-zero component value of 1 
in a coordinate system where the corresponding basis vector for that 
component has length 3 does not have an absolute (physical) length equal to 
1, but to three.

In general, therefore, (see Malvern \cite{Malvern:1969}, for example, for 
further explication), physical components are found from vector components 
via the relations

\begin{equation}
\label{eq23}
v^{\hat {\alpha} }\; = \;v^{\alpha} \sqrt {g_{{\alpha} {\alpha} }}  \quad 
\quad \quad v_{\hat {\alpha} } \; = \;v_{\alpha}  \sqrt {g^{{\alpha} {\alpha 
}}} 
\end{equation}

\noindent
where carets over indices designate physical quantities, and underlining 
implies no summation.

Hence in order to compare theoretical component values with experiment, it 
is necessary to use contravariant components for coordinate differences and 
four-velocities, covariant components for four-momenta, and physical 
components of all component quantities whether covariant or contravariant.

\subsubsection*{4.3.4 Mass-energy of a particle fixed on disk.}

Consider a particle of mass $m$ fixed on the disk at constant $\phi $, $r,$ and 
$z$. Since $d\phi $ = \textit{dr} = \textit{dz} = 0, the four-momentum of the particle in k coordinate 
contravariant components (using metric Eq. (\ref{eq10})) is

\begin{equation}
\label{eq24}
p^{\beta}  = mu^{\beta}  = m{\frac{{dx^{\beta} }}{{d\tau} }} = m{\left[ 
{{\begin{array}{*{20}c}
 {c{\textstyle{{dt} \over {d\tau} }}} \hfill \\
 {0} \hfill \\
 {0} \hfill \\
 {0} \hfill \\
\end{array}} } \right]} = {\frac{{m}}{{\sqrt {1 - {\textstyle{{\omega 
^{2}r^{2}} \over {c^{2}}}}}} }}{\left[ {{\begin{array}{*{20}c}
 {c} \hfill \\
 {0} \hfill \\
 {0} \hfill \\
 {0} \hfill \\
\end{array}} } \right]}
\end{equation}

\noindent
where \textit{dt/d}$\tau $ is found from Eq. (\ref{eq20}).

The mass-energy (non-physical), except for a factor $- c$, and three momenta 
(non-physical) are the four-dimensional conjugate momenta of the d$x^{{\rm 
\beta} }$ and are the components of the covariant four-momentum vector

\begin{equation}
\label{eq25}
\begin{array}{c}
 p_{\alpha}  = g_{\alpha \beta}  p^{\beta}  = {\left[ 
{{\begin{array}{*{20}c}
 { - (1 - {\textstyle{{r^{2}\omega ^{2}} \over {c^{2}}}})} \hfill & {0} 
\hfill & {{\textstyle{{r^{2}\omega}  \over {c}}}} \hfill & {0} \hfill \\
 {0} \hfill & {1} \hfill & {0} \hfill & {0} \hfill \\
 {{\textstyle{{r^{2}\omega}  \over {c}}}} \hfill & {0} \hfill & {r^{2}} 
\hfill & {0} \hfill \\
 {0} \hfill & {0} \hfill & {0} \hfill & {1} \hfill \\
\end{array}} } \right]}{\frac{{mc}}{{\sqrt {1 - {\textstyle{{\omega 
^{2}r^{2}} \over {c^{2}}}}}} }}{\left[ {{\begin{array}{*{20}c}
 {1} \hfill \\
 {0} \hfill \\
 {0} \hfill \\
 {0} \hfill \\
\end{array}} } \right]} \\ 
 \\ 
 = {\frac{{mc}}{{\sqrt {1 - {\textstyle{{\omega ^{2}r^{2}} \over {c^{2}}}}} 
}}}{\left[ {{\begin{array}{*{20}c}
 { - (1 - {\textstyle{{r^{2}\omega ^{2}} \over {c^{2}}}})} \hfill \\
 {0} \hfill \\
 {{\textstyle{{r^{2}\omega}  \over {c}}}} \hfill \\
 {0} \hfill \\
\end{array}} } \right]} \\ 
 \end{array}
\end{equation}

The physical energy $e $of the particle as measured in k is therefore

\begin{equation}
\label{eq26}
\begin{array}{c}
 e\; = \; - p_{\hat {0}} c\; = \; - p_{0} \sqrt { - g^{00}} c\; = \; - p_{0} 
c = \;mc^{2}\sqrt {1 - {\textstyle{{\omega ^{2}r^{2}} \over {c^{2}}}}} \; \\ 
 \\ 
 = \;mc^{2}\; - \;{\textstyle{{1} \over {2}}}m\omega ^{2}r^{2}\; - \;{\rm 
(}{\rm h}{\rm i}{\rm g}{\rm h}{\rm e}{\rm r}\;{\rm o}{\rm r}{\rm d}{\rm 
e}{\rm r}\;{\rm t}{\rm e}{\rm r}{\rm m}{\rm s}{\rm )} \\ 
 \\ 
 = \;mc^{2}\; + \;V_{class} \; + \;\;......\quad . \\ 
 \end{array}
\end{equation}

\noindent
where $V_{{\rm c}{\rm l}{\rm a}{\rm s}{\rm s}}$\textit{} is the classical potential for 
the particle as seen from the rotating frame, and $V_{{\rm c}{\rm l}{\rm 
a}{\rm s}{\rm s}}$\textit{} plus the higher-order terms is the relativistic potential 
energy.

The energy of the particle as seen from K can be found by using the second 
of Eqs. (\ref{eq13}) to transform $p^{{\rm \beta} }$ of Eq. (\ref{eq24}) into the 
four-momentum of the inertial frame $P^{{\rm {\rm B}}}$. By then using 
$G_{{\rm {\rm A}}{\rm {\rm B}}}$, the metric of K, $P_{{\rm A}}$ is found to 
be

\begin{equation}
\label{eq27}
P_{A} = G_{AB} P^{B} = {\left[ {{\begin{array}{*{20}c}
 { - 1} \hfill & {0} \hfill & {0} \hfill & {0} \hfill \\
 {0} \hfill & {1} \hfill & {0} \hfill & {0} \hfill \\
 {0} \hfill & {0} \hfill & {R^{2}} \hfill & {0} \hfill \\
 {0} \hfill & {0} \hfill & {0} \hfill & {1} \hfill \\
\end{array}} } \right]}{\frac{{mc}}{{\sqrt {1 - {\textstyle{{\omega 
^{2}R^{2}} \over {c^{2}}}}}} }}{\left[ {{\begin{array}{*{20}c}
 {1} \hfill \\
 {0} \hfill \\
 {{\textstyle{{\omega}  \over {c}}}} \hfill \\
 {0} \hfill \\
\end{array}} } \right]}
\end{equation}

The mass-energy $E$ of the particle as measured from K is therefore

\begin{equation}
\label{eq28}
\begin{array}{c}
 E\;\; = \;\; - P_{\hat {0}} \,c\;\; = \;\; - P_{0} \sqrt {G^{00}} c\;\;\; = 
\;\; - P_{0} c\;\; = \;\;{\frac{{mc^{2}}}{{\sqrt {1 - {\textstyle{{\omega 
^{2}R^{2}} \over {c^{2}}}}}} }} \\ 
 = \;\;{\frac{{mc^{2}}}{{\sqrt {1 - {\textstyle{{v^{2}} \over {c^{2}}}}} 
}}}\;\; = \;\;mc^{2}\;\; + \;\;{\textstyle{{1} \over {2}}}m\omega 
^{2}R^{2}\;\; + \;\;.......\quad . \\ 
 \end{array}
\end{equation}

\noindent
in full accord with the relativistic mass-energy effect. Note that the total 
energy in both k and K becomes imaginary at $r = c$/$\omega $ where the tangential 
disk speed reaches that of light. That these results were obtained from the 
transformations (\ref{eq8}) (i.e., Eqs. (\ref{eq13}) derived without \textit{ad hoc} Lorentz factors thrown 
in, supports the contention that those transformations are indeed the 
correct ones. 

Note also that $P_{{\rm \Phi} }$\textit{} turns out to be the relativistic angular 
momentum, the conjugate momentum of $\Phi $, as it must be if the 
transformation employed is correct.

\begin{equation}
\label{eq29}
P_{\Phi}  \;\; = \;\;{\frac{{mR^{2}\omega} }{{\sqrt {1 - {\textstyle{{\omega 
^{2}R^{2}} \over {c^{2}}}}}} }}\;\; = \;\;{\frac{{mvR}}{{\sqrt {1 - 
{\textstyle{{v^{2}} \over {c^{2}}}}}} }}
\end{equation}

Further, $p_{{\rm \phi} }$, the relativistic angular momentum as seen from k 
(see Eq. (\ref{eq25})), has the same value as $P_{{\rm \Phi} }$, and is non-zero even 
though the four-velocity component $u^{{\rm \phi} }$ in k is zero.

\section*{5 RAMIFICATIONS OF NON-TIME-ORTHOGONALITY}
\label{sec:mylabel2}

\subsection*{5.1 The Speed of Light}

Consider the line element of Eq. (\ref{eq10}) for a ray of light directed 
tangentially at radius $r$ (with velocity c in K). \textit{dz} =\textit{ dr} = 0, and

\begin{equation}
\label{eq30}
ds^{2} = 0 = \;\; - \;c^{2}(1 - {\textstyle{{\omega ^{2}r^{2}} \over 
{c^{2}}}})dt^{2} + r^{2}d\phi ^{2} + 2r^{2}\omega d\phi dt\quad .
\end{equation}

Solving Eq. (\ref{eq30}) for $d\phi $\textit{} via the standard quadratic equation formula and 
dividing the result by \textit{dt,} one obtains

\begin{equation}
\label{eq31}
{\frac{{d\phi} }{{dt}}}\;\; = \;\; - \;\omega \;\;\pm 
\;\;{\frac{{c}}{{r}}}\quad ,\quad \quad v^{{\rm t}{\rm a}{\rm n}{\rm g}}\;\; 
= \;\;{\frac{{rd\phi} }{{dt}}}\;\; = \;\; - \;r\omega \;\; + \;\;c\;.
\end{equation}

The same result as Eq. (\ref{eq19}).

For velocities in terms of local times on the rim, substitute

\begin{equation}
\label{eq32}
dt = {\frac{{dt_{l}} }{{\sqrt {1 - {\textstyle{{\omega ^{2}r^{2}} \over 
{c^{2}}}}}} }}\quad ,
\end{equation}

\noindent
where \textit{dt}$_{l}$ is time as measured by local standard clocks at $r,$ and hence

\begin{equation}
\label{eq33}
v^{{\rm t}{\rm a}{\rm n}{\rm g}{\rm ,}{\rm p}{\rm h}{\rm y}{\rm s}}\;\; = 
\;\;{\frac{{ - \;r\omega \;\pm \;c}}{{\sqrt {1 - {\textstyle{{\omega 
^{2}r^{2}} \over {c^{2}}}}}} }}
\end{equation}

\noindent
is the exact expression for the first-order approximation Eq. (\ref{eq2}) of the 
Sagnac experiment.

Note that without the off diagonal (non-orthogonal) terms of the metric in 
Eq. (\ref{eq30}) the physical velocity above measured by local standard clocks would 
be $c$. (Delete the last term in Eq. (\ref{eq30}) and substitute the relation Eq. (\ref{eq32}) 
for \textit{dt} in terms of \textit{dt}$_{l}$.)

Fig. 7 helps to explain this effect of non-orthogonality graphically. The 
coordinate axes shown as perpendiculars represent K; the slanted coordinate 
lines represent the local inertial frame at $r$, K$_{{\rm 1}}$; the line MN 
represents the null path of a light ray; and the bold lines represent the 
non-orthogonal coordinate axes for k at the location $r.$ Note that the 
coordinate time axes of k and K$_{{\rm 1}}$ are coincident, as are the 
spatial axes of k and K. Coordinate times $T$ (in K) and $t$ (in k) are equal (see 
dashed horizontal lines representing different values of \textit{cT = ct})$.$ Physical spatial 
distance in K (= $R\Delta \Phi $) and in k (=$ r\Delta \phi $) are equal, 
and we consider $\Delta $ values as small. Coordinate systems for K and 
K$_{{\rm 1}}$ are orthogonal in Minkowski space, the coordinate system for k 
at $r$ is not. In all three systems the path length of MN is zero, since 
pathlength is invariant.

\begin{figure}
\centering
\includegraphics*[bbllx=0.26in,bblly=0.13in,bburx=5.16in,bbury=3.59in,scale=1.00]{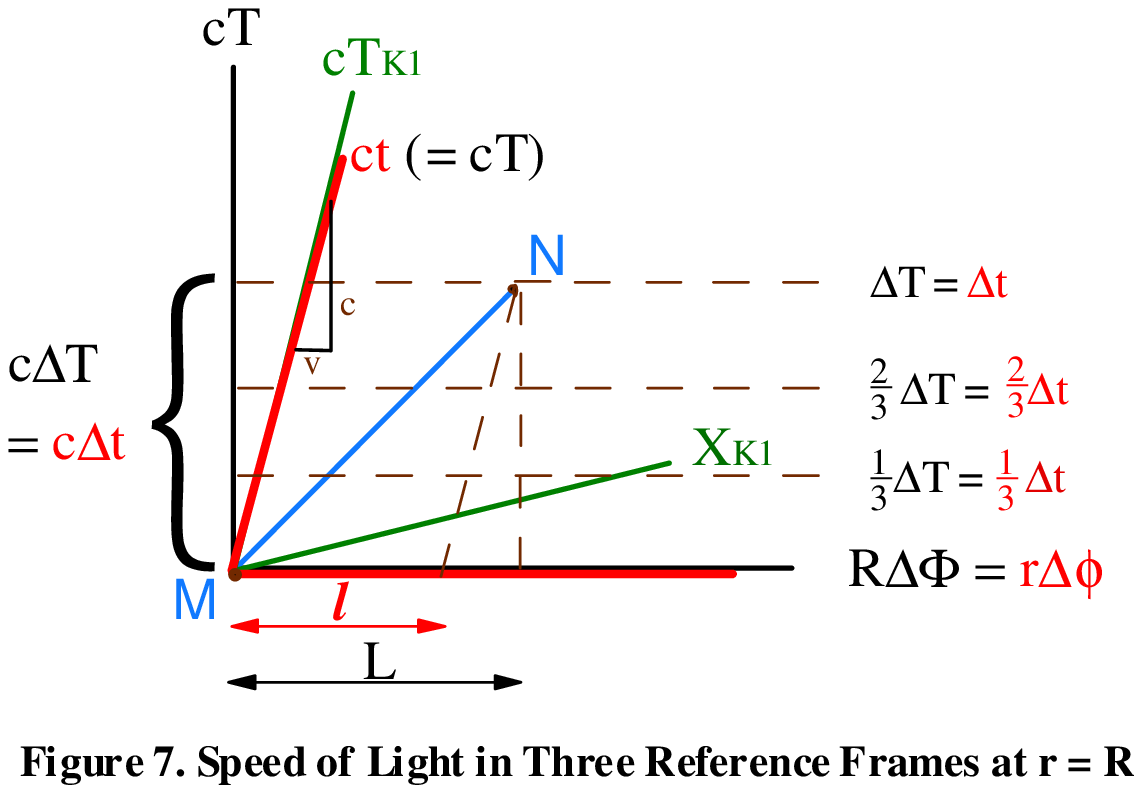}
\end{figure}

\bigskip

Observe that for a given amount of coordinate time (which is the same in 
both k and K, i.e., $c\Delta T$ = $c\Delta t$), the light ray travels a certain 
spatial distance $l $in k, but a greater spatial distance $L$ in K. Hence the speed 
of light measured in k is less than that in K, and this corresponds with the 
plus sign before the $c$ in Eq. (\ref{eq31}). For a light ray in the opposite direction 
(minus sign in Eq. (\ref{eq31})) one can show graphically (with a light ray 
MN$^{{\rm /} }$ in the second quadrant of Fig. 7 at right angles to MN) that 
the corresponding $l$ distance is greater than $L$ and hence the velocity for that 
ray would be greater in k than in K.

Given that the slope of MN is unity, $L = c\Delta T$. Dividing this by $\Delta 
T$, one gets the speed of light in K as $c.$ The k time axis has slope $c/v = c/\omega 
r$, so \textit{l = L - c}$\Delta T$($\omega r/c$). Dividing this by $\Delta T$, one arrives at Eq. 
(\ref{eq31}) for the coordinate speed of light in k (with the plus sign for $c$ since 
light ray MN is traveling in the direction of disk rotation).

Note that in both inertial frames K and K$_{{\rm 1}}$, the speed of light 
ray MN equals c. We can therefore conclude that general relativity principle 
1 remains valid, provided we constrain it to refer to time orthogonal frames 
(such as the Minkowski, Schwarzchild, and Friedman geometries). It does not 
hold for non-time-orthogonal frames.

\subsection*{5.2 The "Surrogate Rods Postulate"}
\label{subsec:mylabel5}

Non-time-orthogonality also reconciles the results of Sec. 2.3 (tangent 
frames can not be used to measure the circumference) with the heretofore 
seemingly universal applicability of the surrogate rods postulate of Sec. 
1.2. We note first, however, that the tangent frames do not, strictly 
speaking, have the same velocity as the disk rim. The rim segment, in 
addition to its linear velocity component $v = \omega r$, has an angular velocity 
$\omega $ which the tangent frame does not. Hence, unlike other successful 
applications of the surrogate rods postulate, the tangent frames here do 
\textit{not} mimic the rim frame velocity in all regards. Therefore, they can not, in 
the truest sense be considered "co-moving" as many prior researchers have 
assumed. To see the effect of this in terms of non-time-orthogonality, we 
first consider the underlying principles on which the surrogate rods 
postulate is based.

Fig. 8 shows two spatially coincident standard rods with zero relative 
velocity, the first of which is fixed in a rectilinearly accelerating frame 
k$_{{\rm a}}$, and the second of which is fixed in an inertial frame K. 
Consider two light flash events A and B located at the endpoints of both 
rods. An observer in K at the centerpoint of the inertial rod sees both 
flashes at the same time and concludes they were simultaneous as seen from 
the K frame. Likewise, an observer on the k$_{{\rm a}}$ rod halfway between 
A and B would see them at the same time as well and know that the events 
were also simultaneous as seen from the k$_{{\rm a}}$ frame. That is, 
$\Delta t_{a}$ = $\Delta T$ = 0 between A and B. Since the proper spacetime 
length between the two events is the same as seen from both frames (i.e.,\textit{} $\Delta 
s_{a}$ = $\Delta S$), then the spatial length between them must also be 
equal, and the length measured by rods in k$_{{\rm a}}$ and K between A and 
B are equal. Similar arguments can be made for acceleration in the rod 
lengthwise direction, as well as for gravitational body forces induced by a 
massive body.

\begin{figure}
\centering
\includegraphics*[bbllx=0.26in,bblly=0.13in,bburx=4.44in,bbury=2.07in,scale=1.00]{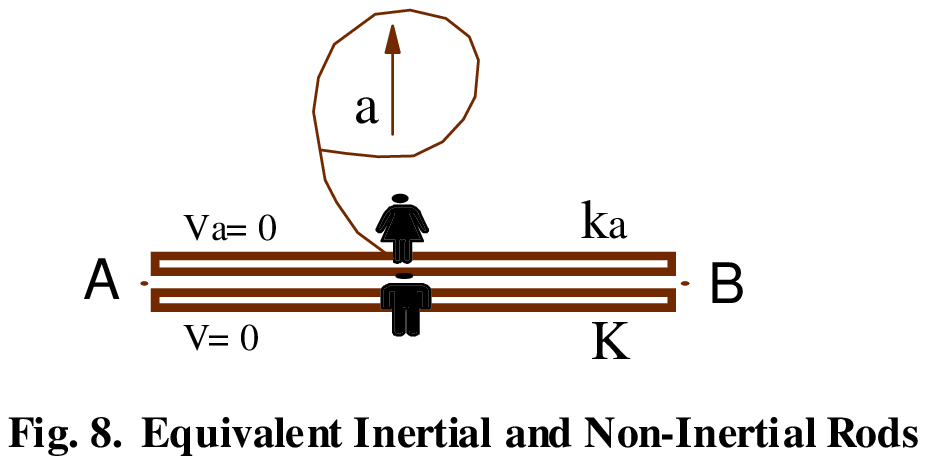}
\end{figure}

\bigskip

Hence the surrogate rods postulate is merely a restatement of the proper 
spacetime path length invariance postulate for the special case where 
$\Delta $(time) = 0 for both observers (which it is for zero relative 
velocity and time orthogonal frames). Figs. 9 and 10 reveal what this means 
in the context of non-time-orthogonal reference frames.

Fig. 9 depicts two inertial reference frames, K and K$_{{\rm 1}}$, in 
relative motion, and serves as a review of the cause of the Lorentz 
contraction effect. A rod fixed in K$_{{\rm 1}}$ has length$ L_{{\rm 1}}$ =$ L$ as 
seen from K$_{{\rm 1}}$. The endpoints (and all points between) of$ L_{{\rm 
1}}$ move with velocity $v $relative to K along world lines parallel to the 
K$_{{\rm 1}}$ time axis. Lorentz contraction arises because the observer in 
K sees rod endpoint events as A and C, whereas the observer in K$_{{\rm 1}}$ 
sees them as A and B. $ L_{{\rm 1}}$', the distance between A and C is less 
than$ L$, the distance between A and B, by the Lorentz contraction factor. 
Though it is beside the point we are in the process of making, the Lorentz 
contraction effect is thus seen to be little more than an optical illusion 
fostered on us by lack of agreement in simultaneity. No Lorentz contracted 
object ever feels compressed.

\begin{figure}
\centering
\includegraphics*[bbllx=0.26in,bblly=0.13in,bburx=5.34in,bbury=2.50in,scale=1.00]{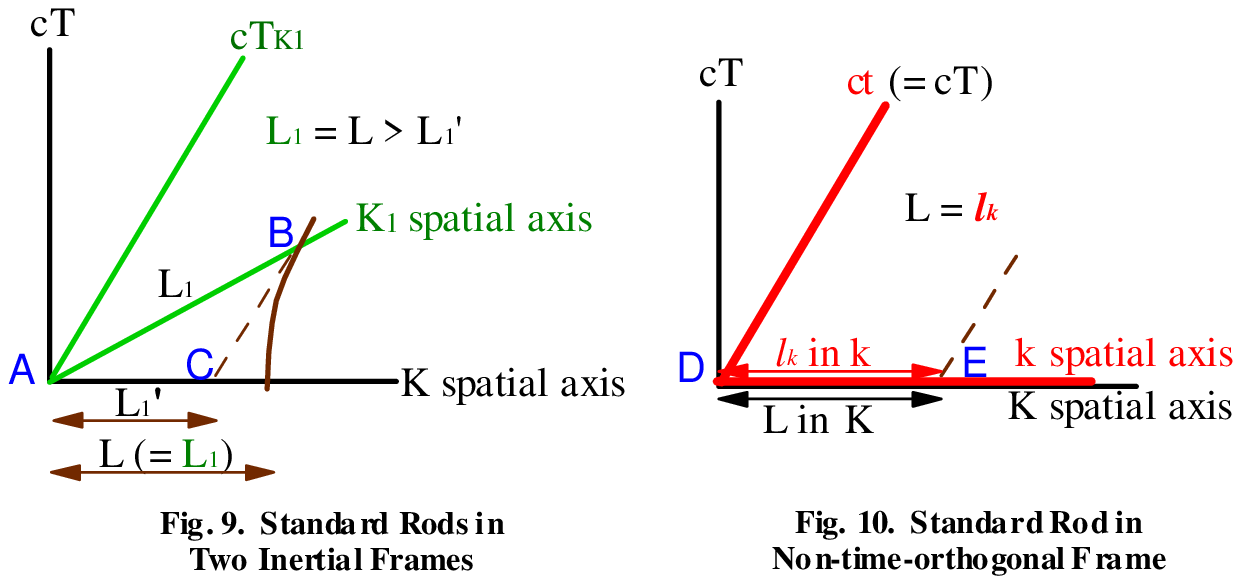}
\end{figure}

\bigskip

In contrast with Fig. 9, Fig. 10 shows the rotating coordinate frame k at 
radius $r$ superimposed with the non-rotating inertial frame K. In Fig. 10 we 
show two circumferentially aligned standard rods, one designated by 
$l_{k}$ which rides with the disk and the other, $L$, fixed in K. When both rods 
are at rest in the same inertial system, they have equal length, i.e., 
$l_{k}$ = $L$. When the disk is spinning they also have equal length, as seen by 
both k and K observers, since, as discussed earlier, the same endpoint 
events (D and E) of both rods are seen as simultaneous in both frames. Yet, 
due to the special nature of the non-time-orthogonal frame k at $r$, the 
$l_{k}$\textit{} rod has a non-zero velocity relative to the $L$ rod. Every point on the 
$l_{k}$ rod in Fig. 10 moves with the same velocity as every point on the 
$L_{{\rm 1}}$ rod in Fig. 9. Yet $L_{{\rm 1}}$ looks contracted from K, 
whereas $l_{k}$ does not, i.e.,

\begin{equation}
\label{eq34}
L_{\;1} {\rm}  = {\rm} \;L{\rm}  = {\rm} {\rm} l_{k} {\rm} {\rm}  > {\rm 
}{\rm} {L}'_{\;1} 
\end{equation}

That is, even though $L_{{\rm 1}}$ and $l_{k}$ have the same velocity as seen 
from K, they do not have the same spatial length as seen from K. Prior 
researchers have almost universally assumed they do.

That $L_{{\rm 1}}$ looks contracted as seen from k can be corroborated by 
superimposing the k frame of Fig. 10 with the K$_{{\rm 1}}$ frame of Fig. 9. 
Hence, two rods with the same velocity, one in a time orthogonal frame and 
one in a non-time-orthogonal frame do not have equal lengths. We conclude 
that the surrogate rods postulate is only valid for time orthogonal frames.

\subsection*{5.3 The Large Radius, Small $\omega $ Limit}

Several researchers [13,18] have considered the limiting case of very large 
radius, small angular velocity, with large circumferential velocity $v = \omega 
r. $ In this case acceleration $v^{{\rm 2}}/r$ approaches zero, and it is argued 
there is no way to discern between such a frame and an inertial frame. 
Hence, a circumferential segment of the rotating frame in such case must 
approximate an inertial (Lorentzian) frame. 

The answer to this conundrum lies in non-time-orthogonality. From Fig. 7 it 
can be seen that the slope of the time axis in the k frame is c/$v$, and as we 
have shown, the Sagnac effect, the lack of Lorentz contraction, and all 
other peculiarities of the rotating frame are derivable from 
non-time-orthogonality, i.e., the slope of that axis. But in taking the 
limit described, $v$ remains constant, and hence so does the slope of the time 
axis. Therefore, all of the non-Lorentzian phenomena heretofore described 
for rotating frames are unmitigated in passing to the limit. (See also Sec. 
6.)

Contrary to what many claim, an observer on this limiting case frame can 
determine she is rotating. In fact, three experiments can reveal this. The 
first is the Sagnac experiment. The second is described in Section 6.2 
below. The third involves measuring the mass of a known entity such as an 
electron which varies relativistically with the potential energy, i.e., as a 
function of $v^{{\rm 2}}$ alone as in Eq. (\ref{eq26}), and hence one can readily 
determine $v.$

\section*{6 THE NEW THEORY AND EXPERIMENT}

\subsection*{6.1 Michelson-Morley Revisited}

Given the speed of our planet around its sun, and the speed of our solar 
system around its galactic center, one might ask why measurements on our 
planet (which could be considered as part of a frame rotating about the 
center of each of these systems) do not seem to exhibit the aforementioned 
non-Lorentzian properties. In particular, why did Michelson and Morley not 
find the speed of light in the direction of galactic rotation different from 
that in other directions? Note that given the Sagnac results, this question 
has an empirical imperative which is independent of any theory, i.e., any 
particular rotating frame analysis.

The answer, the author submits, is that bodies in gravitational orbits 
follow geodesics, i.e., they are in "free fall". That is, they are in 
locally inertial frames and therefore obey Lorentzian mechanics. Objects 
fixed in "true" rotational frames, on the other hand, are held in place by 
non-gravitational forces, do not travel geodesic paths, and exhibit Sagnac 
type characteristics. Hence, the only effective rotational velocity for the 
earth is the earth surface velocity about its own (inertial) axis. Michelson 
and Gale \cite{Michelson:1925} did in fact measure the Sagnac effect for 
the earth's surface velocity in the 1920's. And in order to be maintain 
accuracy, the Global Positioning System must apply a Sagnac velocity 
correction to its electromagnetic signals \cite{Allan:1985}.

\subsection*{6.2 Modern Michelson-Morley Experiments}

The most significant experiment, however, and the most accurate 
Michelson-Morley type test to date is that of Brillet and Hall 
\cite{Brillet:1979}. They found a "null" effect at the $\Delta t$/$t $= 
3X10$^{{\rm -} {\rm 1}{\rm 5}}$ level, ostensibly verifying standard 
relativity theory to high order. However, to obtain this result they 
subtracted out a persistent ``spurious'' signal of amplitude 2X10$^{{\rm - 
}{\rm 1}{\rm 3}}$ at twice the apparatus rotation frequency. 

Compare this anomalous signal to that predicted by the presently proposed 
theory. The velocity of the earth surface at 40$^{{\rm o}}$ latitude, where 
the Brillet and Hall experiment was performed, is .355 km/sec. If the speed 
of light is truly increased or decreased by this amount in the direction of 
rotation, then a Michelson-Morley experiment (see Fig. 11) with one leg in 
the direction of the velocity would yield \cite{Born:1965}

\begin{equation}
\label{eq35}
{\frac{{\Delta t}}{{t}}}\;\; \cong \;\;{\textstyle{{1} \over 
{2}}}{\frac{{v^{2}}}{{c^{2}}}}
\end{equation}

\noindent
where $t $is the round-trip time for one leg and $\Delta t$ is the difference in 
time taken between the leg aligned with the velocity vector and the leg 
perpendicular to that vector.

\begin{figure}
\centering
\includegraphics*[bbllx=0.26in,bblly=0.13in,bburx=5.44in,bbury=2.48in,scale=1.00]{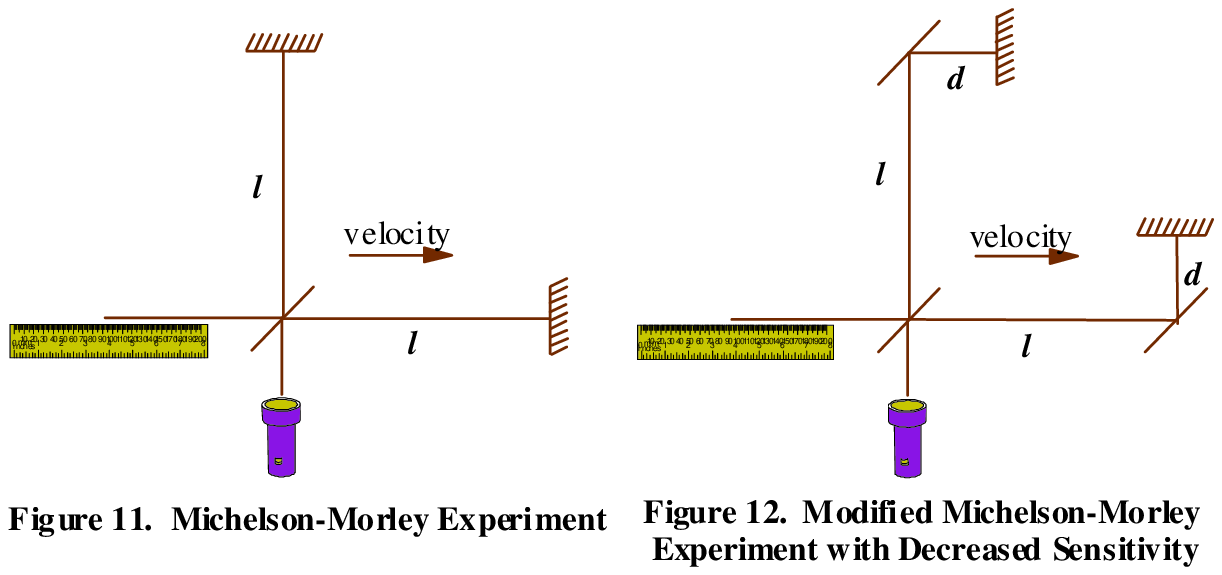}
\end{figure}

\bigskip

Brillet and Hall rotated their equipment about a vertical axis (always 
perpendicular to the earth surface velocity). In such a case Eq. (\ref{eq35}) would 
yield the peak-to-peak amplitude of the signal. To compare with Brillet and 
Hall's reported single peak amplitude signal, one must then divide Eq. (\ref{eq35}) 
by two. Using .355 km/sec in Eq. (\ref{eq35}) and dividing by two, one gets $\Delta 
t$/$t $= 3.5X10$^{{\rm -} {\rm 1}{\rm 3}}$. Fig. 12 can help to explain the 
reason for the discrepancy in this number and the reported value.

In Fig. 12 the two light rays no longer travel solely on two perpendicular 
paths. In the Brillet and Hall experimental apparatus a similar 
configuration to that of Fig. 12 was used, apparently to accommodate the 
laser equipment which provided such extraordinary accuracy. (See Fig. 1 in 
Brillet and Hall.) Note that if the two shorter legs $d_{{\rm} }$were each 
25\% the length of the primary legs $l$, then the time $t$ in Eq. (\ref{eq35}) would be 
1.25 times that of Fig. 11. Note also that the lower of the $d$ legs is aligned 
perpendicularly to the velocity, whereas the lower path is intended to 
monitor light speed in the direction of the velocity vector. If both it and 
the upper $d$ leg were 25\% of the primary legs in length, one could therefore 
expect a 25\% reduction in $\Delta t$ as well. Hence the total signal $\Delta 
t/t $would be a reduced by a factor of .75/1.25 = 60\% .

Although Brillet and Hall did not provide the pertinent dimensions in their 
paper, from the sketch of their equipment one can conclude that the 25\% 
figure used above may be fairly accurate. To account for this approximation 
we can assume a signal strength modification factor of 50\% to 70\% . For 
these percentages, the expected signal range is 1.7-2.5X10$^{{\rm -} {\rm 
1}{\rm 3}}$, in remarkable agreement with the measured value of 2X10$^{{\rm 
-} {\rm 1}{\rm 3}}$.

Note also that in the Brillet and Hall experiment signals were exchanged 
electronically between the two legs, and that actual fringing at a single 
location was not measured. Hence, we may expect the variation in light speed 
to affect these transmission signals as well, introducing additional error. 

The author also investigated possible mitigating effects from "frame 
dragging" (a $t\phi $ off diagonal term appearing in the metric due to the 
earth's angular momentum), and the chord path effect (the light ray parallel 
to the velocity actually travels a chord of the arc length, not the arc 
length itself). These were found to be negligible to orders of magnitude 
well beyond 10$^{{\rm -} {\rm 1}{\rm 3}}$.

For completeness, we also mention the results of Hils and Hall 
\cite{Hils:1990} which found no variations to an order of 2X10$^{{\rm - 
}{\rm 1}{\rm 3}}$ (only coincidentally the same number as above.) However, 
the Hils and Hall apparatus was fixed relative to the earth's surface and 
hence was immune to variations of the type predicted by the present theory 
(which would manifest only if the apparatus were rotated relative to the 
earth surface.)

\subsection*{6.3 A Proposed Experiment}
\label{subsec:mylabel6}

We propose another experiment using a combined Sagnac and Michelson-Morley 
type apparatus in order to further test the present theory. For this 
experiment light is emitted from a rotating disk center in the manner of the 
Sagnac experiment, but a Michelson-Morley type apparatus is mounted on the 
disk rim. When the light from the center reaches the rim it is split into 
two components. One of these travels along the circumference a short 
distance and then is reflected directly backwards (rather than further out 
around the rim.) The other component of the light is reflected in the z 
direction (perpendicular to the disk) an equal distance and reflected 
directly backwards as well. The two returning components are then deflected 
toward the disk center where fringe effects are measured. Accelerating the 
disk, and accounting for elastic deformation, one should find the degree of 
fringing varies. The standard theory predicts no such variation.

\section*{7 ELECTRODYNAMICS, MECHANICS, AND SPACETIME}

With regard to electrodynamics, Ridgely [33] has recently used covariant 
constitutive equations in an elegant analysis to answer a troubling question 
cogently posed by Pellegrini and Swift [32]. He uses transformation (\ref{eq8}) to 
derive electrodynamic results for the rotating frame k, not the tangent 
frames K$_{{\rm i}}$, and finds that those results match what one would find 
by simply applying Maxwell's equations and traditional special relativity to 
the tangent frames.

We can conclude the following. Only with the theoretical approach shown 
herein can one obtain internally consistent results which agree with all 
experiments. However, for the purposes of mass-energy, momentum, and time 
dilation calculations (shown herein) and Maxwell's equations (shown by 
Ridgely), one can get away with assuming that the tangent frames represent 
the rotating frame and using traditional special relativity. That is, in 
these cases the laws of nature conspire to make both the present and the 
traditional analysis produce the same result for observers in K (i.e., 
mass-energy dependence on $\omega $r, electric polarization, etc.). When it 
comes to matters of time (synchronization, simultaneity), space (curvature), 
and 

Michelson-Morley/Sagnac type experiments, however, then analysis must be 
confined to the rotating frame itself, else the inconsistencies of Sec. 2 
and inexplicable ``spurious'' experimental signals inevitably arise.

It therefore appears that the rotating disk problem may have, at long last, 
been completely solved. According to Ridgely's results and the theory 
proposed herein, no paradoxes remain and all theory matches up with the 
physical world as we know it. 

\section*{8 SUMMARY AND CONCLUSIONS}

\subsection*{8.1 New Theory Predictions}

The lack of Lorentz contraction, agreement in simultaneity, flatness of the 
disk surface, non-invariant/non-isotropic speed of light, and time dilation 
can all be derived in two different ways: (i) directly from transformation 
(\ref{eq8}); and (ii) from the Sagnac experiment. All but the asterisked phenomena 
summarized below can be determined in at least two ways (i.e., 
transformation (\ref{eq8}), experiment, or thought experiment based on Sagnac). An 
asterisk (*) indicates the conclusion depends only on the validity of 
transformation (\ref{eq8}).

\bigskip

1. The subspace surface of a rotating disk is flat.

2. No circumferential Lorentz contraction exists. Further, no 
relativistically induced tensile stress develops as there is no kinematic 
imperative for the disk circumference to try to Lorentz contract.

3. Observers anywhere in the rotating frame and observers in the 
non-rotating frame all agree on simultaneity.

4. Velocities (angular and translational*) add directly frame-to-frame and 
not relativistically.

5. Angular velocities are absolute and have no upper speed limitation.

6. Rods in inertial frames with velocities equal to the tangent velocities 
at a given disk radius can not be used to measure the circumference, since

\noindent
a. the "surrogate frames postulate" for equivalence of inertial and 
non-inertial standard rods is not valid for the rotating frame, and is 
generally invalid for any non-time-orthogonal frame, and

\noindent
b. doing so leads to a discontinuity in time.

7. Light has a null path length, yet the local speed of light in the 
rotating frame (and all non-time-orthogonal frames) is not isotropic and 
generally not equal to $c$.

8. Time dilation does occur, but it is not symmetric, i.e., rotating and 
non-rotating observers agree that time dilation occurs on the disk relative 
to the stationary frame.

9. A particle fixed on the disk exhibits relativistic mass-energy dependence 
on tangential velocity. (No asterisk since cyclotron experiments validate 
this effect.)

10. Only the theory proposed herein yields self consistent results which 
completely conform with physical reality. However, use of tangent frames and 
traditional special relativity produce the same results for a certain subset 
of phenomena.

\subsection*{8.2 Comparison of Various Perspectives}

The proposed theory resolves all difficulties with the traditional disk 
analysis delineated in Sec. 2. 

Both the proposed theory and the traditional analysis agree with cyclotron 
experiments, i.e., they both predict time dilation (longer particle decay 
times), as well as relativistic mass-energy dependence on speed. Both 
theories are also consonant with the Phipps \cite{Phipps:1974} experiment 
which has been used to discount certain other prior approaches to the 
problem not based on transformation (\ref{eq8}). Importantly, however, the new 
theory predicts the results of the Brillet and Hall [38] experiment which 
the standard theory does not.

The new theory also agrees with part of Gr{\o}n's first work \cite{Ref:8} 
on the standard approach, where he uses transformation (\ref{eq8}) and concludes 
from it that simultaneity on the disk and in the lab are the same. However 
his latter paper [10] employs tangent frames analysis and recounts purported 
difficulties in accelerating the disk which are the direct result of 
disagreement in simultaneity. In the presently proposed theory no such 
kinematic restriction on disk acceleration exists, and there is no mechanism 
by which any such disk would rupture from relativistically induced tensile 
stress, as has been contended by Einstein, Gr{\o}n, and others.

Table I summarizes the similarities and differences between the approaches 
of Einstein, Gr{\o}n, and the proposed theory.

\bigskip

\begin{center}
\textbf{TABLE I. COMPARISON OF VARIOUS DISK ANALYSES}
\end{center}

\bigskip

\newcommand{\PreserveBackslash}[1]{\let\temp=\\#1\let\\=\temp}
\let\PBS=\PreserveBackslash
\begin{longtable}
{|p{179pt}|p{61pt}|p{54pt}|p{54pt}|}
a & a & a & a  \kill
\hline
& 
\underline {Einstein}& 
\underline {Gr{\o}n}& 
\underline {ThisPaper} \\
\hline
Postulates agree with experiment?& 
No& 
No& 
Yes \\
\hline
Discontinuity in time?& 
Yes& 
Yes \& No& 
No \\
\hline
Clock synchronized with itself?& 
No& 
No& 
Yes \\
\hline
Closed paths allowable?& 
No& 
No& 
Yes \\
\hline
"Tension" in time required?& 
Yes& 
Yes& 
No \\
\hline
Predicts Brillet \& Hall anomaly?& 
No& 
No& 
Yes \\
\hline
Agrees with cyclotron experiments?& 
Yes& 
Yes& 
Yes \\
\hline
Relativistic mass-energy?& 
Yes& 
Yes& 
Yes \\
\hline
Time dilation on disk?& 
Yes& 
Yes& 
Yes \\
\hline
Lorentz contraction effect?& 
Yes& 
Yes& 
No \\
\hline
Disk surface is curved?& 
Yes& 
Yes& 
No \\
\hline
Relativistically induced disk stress?& 
Yes& 
Yes& 
No \\
\hline
Same simultaneity: disk and lab?& 
No& 
Yes \& No& 
Yes \\
\hline
Time as defined is observable?*& 
No& 
No& 
Yes \\
\hline
Transformation is Galilean type?& 
No& 
Yes& 
Yes \\
\hline
Restricts surrogate rods principle?& 
No& 
No& 
Yes \\
\hline
Speed of light = $c$ on disk?& 
Yes& 
No$^{{\rm \dag} }$& 
No \\
\hline
Agrees with Phipps experiment?& 
Not treated& 
Yes& 
Yes \\
\hline
\end{longtable}

\bigskip

*See Appendix 

$^{{\rm \dag} }$ Result derived from transformation (\ref{eq8}), but effect on 
relativity postulates not considered.

\bigskip

\textbf{ACKNOWLEDGMENT.} The author would like to express his sincerest 
gratitude to Robin Ticciati and Arthur Swift for reading early versions of 
the manuscript, for offering valuable suggestions, and above all, for their 
encouragement and support for this work, despite its paradigm-challenging 
nature.

\bigskip

\section{APPENDIX: THE METRIC OF PRIOR TREATMENTS}

Landau and Lifshitz [12], M{\o}ller [11], Strauss [9], and Gr{\o}n [10] have 
all discussed a three dimensional submetric of the four-dimensional rotating 
frame metric defined by

\[
\begin{array}{l}
 \\ 
 \quad \quad \quad \quad \quad \quad \quad \quad \gamma _{ij} \;\; = 
\;\;g_{ij} \;\; - \;\;{\frac{{g_{0i} g_{0j}} }{{g_{00}} }}\quad .\quad \quad 
\quad \quad \quad \quad \quad \quad \quad {\rm (}{\rm A}{\rm 1}{\rm )} \\ 
 \\ 
 \end{array}
\]

So defined, $\gamma _{ij} $ represents the spatial metric which is locally 
orthogonal to the local proper time axis. Some of these authors have then 
used this metric to determine whether the space of the rotating disk is flat 
or not, and have concluded that it is curved. In fact, using the metric of 
Eq. (\ref{eq11}) for $g_{{\rm \alpha} {\rm \beta} }$ in Eq. (A1) one finds

\[
\begin{array}{l}
 \\ 
 \quad \quad \quad \quad \quad \quad \quad \quad \gamma _{ij} \;\; = 
\;\;{\left[ {{\begin{array}{*{20}c}
 {1} \hfill & {0} \hfill & {0} \hfill \\
 {0} \hfill & {{\frac{{r^{2}}}{{1 - {\textstyle{{r^{2}\omega ^{2}} \over 
{c^{2}}}}}}}} \hfill & {0} \hfill \\
 {0} \hfill & {0} \hfill & {1} \hfill \\
\end{array}} } \right]}\quad \quad \quad \quad \quad \quad \quad \quad \quad 
\quad {\rm (}{\rm A}{\rm 2}{\rm )} \\ 
 \\ 
 \end{array}
\]

A line element around the circumference then becomes

\[
\begin{array}{l}
 \\ 
 \quad \quad \quad \quad \quad \quad \quad ds\;\; = \;\;{\frac{{r}}{{\sqrt 
{1 - {\textstyle{{r^{2}\omega ^{2}} \over {c^{2}}}}}} }}d\phi \;\; = 
\;\;{\frac{{r}}{{\sqrt {1 - {\textstyle{{v^{2}} \over {c^{2}}}}}} }}d\phi 
\quad \quad \;\;\quad \quad \quad \quad {\rm (}{\rm A}{\rm 3}{\rm )} \\ 
 \\ 
 \end{array}
\]

From Eq. (A3) it is obvious that the circumference $C $is not equal to 2$\pi 
r, $and in fact is equal to what Einstein and some of the above authors have 
claimed. 

Further, Riemann for the metric of Eq. (A2)\textit{} is non-zero$.$

However, the metric Eq. (A2) is derived for a differential line element 
having simultaneous starting and ending points as measured by local inertial 
clocks. In other words it assumes that a local inertial frame is aligned 
with \textit{ds} and that measurement is carried out such that the endpoints of \textit{ds} are 
simultaneous in that inertial frame. But this is nothing other than the type 
of integration path we investigated herein with the aid of Fig. 3\textit{} (solid 
line).\textit{} As demonstrated, such an integration cannot be carried out around a 
closed\textit{} path on the surface of the disk wherein the starting and ending points 
are simultaneous, and hence, it is meaningless as a physical measure of the 
disk circumference. 

Adler, Bazin, and Schiffer \cite{Adler:1975} use the time transformation

\[
\begin{array}{l}
 \\ 
 \quad \quad \quad \quad \quad \quad \quad \;\;dt\ast \;\; = \;\;dt\;\; - 
\;\;{\frac{{\omega r^{2}}}{{c^{2} - \omega ^{2}r^{2}}}}d\phi \quad \quad 
\quad \quad \quad \quad \quad \quad \;{\rm (}{\rm A}{\rm 4}{\rm )} \\ 
 \\ 
 \end{array}
\]

\noindent
in Eq. (\ref{eq10}) and obtain the same metric Eq. (A2), where the new coordinate 
time is then $t$*.

However, the transformation Eq. (A4) is like any transformation in that it 
effectively shifts one to a different reference frame. Hence the time $t$* as 
defined no longer represents time on the rotating frame itself, but some 
other time. This other time definition, we contend, has no meaning in the 
sense of being actually observable in the physical world by any possible 
observer. In essence, it represents an observer who miraculously skips from 
tangent inertial frame to tangent inertial frame without the concomitant 
acceleration and rotation associated with the disk itself. Not only is this 
not possible, but such a definition of time leads to a temporal 
discontinuity, as we have shown.

\bigskip

\textbf{REFERENCES AND NOTES}


\begin{thebibliography}{27}
\bibitem{This:1} This is strictly true only, as we shall see, for time orthogonal frames. For such frames this is equivalent to saying light always travels null paths.
\bibitem{Ref:6} Ref. --19, pp. 173. Using the metric for rectilinearly accelerating systems one can immediately see that time is dilated along the direction of the acceleration. Calculation of the Riemann tensor from the metric results in zero value for Riemann for the 4D space of the accelerating system. Since the metric is diagonal, any subspace of two or more coordinates is also Riemann flat.
\bibitem{There:1} There is a little subtlety here. Calculation of the Lorentz contraction actually involves rod endpoints which appear to be different events to different observers. The example assumes the same two events are measured by both observers. However, the conclusion remains valid. If the endpoint events look simultaneous to two different observers, then the rod length measured must be the same for each.
\end{thebibliography}
\end{document}